\documentclass[twocolumn]{aastex631}

\graphicspath{{./}{figures/}}

\usepackage{amsmath,amssymb,mathrsfs,graphicx,float,latexsym,url}
\usepackage{epstopdf,ragged2e}
\usepackage{natbib,url}
\usepackage{hyperref}
\usepackage{ulem}

\newcommand{\bq}{\boldsymbol{q}}

\newcommand{\tr}{\tilde{r}}

\newcommand{\Mc}{M_{\text{c}}}
\newcommand{\Rc}{R_{\text{c}}}
\newcommand{\Ms}{M_{\star}}

\newcommand{\bvc}{\boldsymbol{v}_{\text{c}}}
\newcommand{\bvs}{\boldsymbol{v}_{\star}}

\newcommand{\htwo}{$\text{H}_{2}$}

\def\apj{{ApJ}}
\def\aj{{AJ}}

\def\apjl{{ApJL}}

\def\aap{{A\&A}}

\def\araa{{Annual Review of Astronomy and Astrophysics}}

\def\mnras{{MNRAS}}

\def\nat{{Nature}}

\def\04a{{2004 a}}
\def\04b{{2004 b}}

\def\gta{\ifmmode\stackrel{>}{_{\sim}}\else$\stackrel{>}{_{\sim}}$\fi}
\def\lta{\ifmmode\stackrel{<}{_{\sim}}\else$\stackrel{<}{_{\sim}}$\fi}

\begin{document}

\title{Tidal disruption of ``snow clouds'' by unassociated stars}

\correspondingauthor{Mark Walker}
\email{arthur.suvorov@manlyastrophysics.org, mark.walker@manlyastrophysics.org}

\author[0000-0002-3112-5004]{Arthur G. Suvorov}
\affiliation{Manly Astrophysics, 15/41-42 East Esplanade, Manly, NSW 2095, Australia}
\affiliation{Departament de F{\'i}sica Aplicada, Universitat d'Alacant, Ap. Correus 99, E-03080 Alacant, Spain}

\author[0000-0002-5603-3982]{Mark A. Walker}
\affiliation{Manly Astrophysics, 15/41-42 East Esplanade, Manly, NSW 2095, Australia}

\begin{abstract}
\noindent It has been suggested that star-forming galaxies may host a substantial, dark reservoir of gas in the form of planetary-mass molecular clouds that are so cold that \htwo\ can condense. Here we investigate the process of tidal disruption of such ``snow clouds'' by close passage of field stars. We construct a suite of simulations using the hydrodynamic formalism introduced by Carter and Luminet, and use it to explore the properties of the resulting tidal debris. The debris streams are tiny structures that are highly over-pressured relative to the ambient ISM. They are also unusual in their composition --- initially consisting of cold, gaseous He together with \htwo\ ``snowballs'' that may be as much as a metre in size. Each stream expands and cools and is subsequently shocked as it ploughs through the ISM; the snowballs are gradually eroded by the shocked gas. Snowballs streaming through the shocked ISM create microstructured plasma that is somewhat reminiscent of the ``scattering screens'' revealed by radio-wave scintillation studies. However, the tidal disruption rate is too low to account for the observed number of scattering screens if, as we assume here, the stars and clouds have no prior physical association so that disruptions occur as a result of chance encounters between stars and clouds.
\end{abstract}



\section{Introduction}\label{sec:intro}
It is a well-established idea that a reservoir of dark gas may help to explain the observed properties of star-forming galaxies \citep{pfen94}, and many studies have been undertaken with a view to constraining, or uncovering, interstellar gas in hard-to-see forms  \citep[e.g.][]{2005Sci...307.1292G,2011A&A...536A..19P,2014A&A...561A.122L,2015AJ....149..123A,2018ApJ...862..131M, 2020A&A...643A.141M, 2023A&A...670A.115S}. As originally pointed out by \cite{pc94}, gas clouds that are simultaneously cold and dense will lie close to the \htwo\ phase equilibrium curve, and it is therefore likely that they contain hydrogen in condensed form: the hypothesised reservoir ought to be made up of ``snow clouds''.

Despite being a material that has been extensively studied in the laboratory \citep[e.g.][]{1980RvMP...52..393S}, solid \htwo\ in the interstellar environment has not been much explored. That is because the pure form of the solid is expected to sublimate rapidly \citep{1969MNRAS.144..411F,1969Natur.224..251G}, and was therefore thought to be of no practical relevance. However, recent developments have changed the outlook for solid \htwo. First, in the ISM the solid is expected to charge-up; the resulting electric field increases the binding energy of the molecules, and the sublimation rate can be many orders of magnitude below that of the pure solid \citep{2013MNRAS.434.2814W}. Secondly, quantum mechanical calculations have demonstrated that solid \htwo\ is expected to be rich in spectral features: it has been suggested to be the carrier of the major mid-infrared bands of the ISM \citep{2011ApJ...736...91L}, and it has also been suggested to be the carrier of the many optical absorption lines known as the Diffuse Interstellar Bands \citep{2022ApJ...932....4W}. And thirdly, it has been recognised that the interstellar object 1I/`Oumuamua \citep{2017Natur.552..378M} might be principally composed of solid \htwo\ \citep{2018A&A...613A..64F, 2020ApJ...896L...8S}.

These developments motivate theoretical exploration of the possible astrophysical manifestations of both solid \htwo\ and the sort of clouds within which \htwo\ can condense. Structural models of spherically symmetric \htwo\ snow clouds yield masses in the planetary range, and radii comparable to those of planetary orbits \citep{ww19}. Thus although the model clouds are well-defined entities, with a hard outer edge, they do have much lower volume-/column-densities than planets and stars, and that makes them vulnerable to collisions.

At the hypersonic speeds that are expected inside galaxy halos, physical collisions are destructive and clouds that have a low column-density will almost all be wiped out over the course of a Hubble time \citep{1996ApJ...472...34G}. That constraint was investigated by \cite{1999MNRAS.308..551W}, who found that a dark halo initially made entirely of snow clouds evolves under the influence of collisions to yield a specific, power-law relationship between the visible mass content and rotation speed of a galaxy. Using published data, \cite{1999MNRAS.308..551W} further demonstrated that real galaxies do indeed follow the predicted relationship, thus providing a physical basis for the Tully-Fisher relation\footnote{It is now widely accepted that a relationship between visible mass and circular speed does indeed underlie the Tully-Fisher relation. Ironically, the name ``Baryonic Tully-Fisher Relation'' is now commonly used for the visible-mass:circular-speed relation, following \cite{2000ApJ...533L..99M}.} of star-forming galaxies. Matching the model to data allowed \cite{1999MNRAS.308..551W} to determine the mean column-density of the individual clouds: $\langle \Sigma\rangle \equiv \Mc / \pi \Rc^2 \simeq 140\;{\rm g \, cm^{-2}}$, where $\Mc$ and $\Rc$ denote a typical cloud mass and radius, respectively.

If snow clouds and stars are both present then another two-body process comes into play: tidal disruption of clouds by stars. This process is expected to occur with a much larger cross-section than physical collisions between these two species (see \S\ref{sec:colr}). Tidal disruption events (TDEs) provide a second, separate pathway to destruction for snow clouds --- a pathway that becomes increasingly important as the number-density of stars increases. We therefore have the possibility of interstellar gas in potentially unusual kinematic and morphological forms.

As one possible example of the latter: \citet{wang21} discovered five, rapidly scintillating radio sources all in a line. They inferred that those sources are seen through a long, thin plasma filament, and suggested that the filament could have arisen as the result of tidal disruption of a snow cloud by a star. 

In respect of the kinematics of the debris we are interested in both the individual characteristics and the ensemble of all collision products, because 21cm line observations constrain the amount of atomic gas that is moving at high speeds relative to the local standard of rest (LSR) \citep{1990ARA&A..28..215D,wakk97}. It is possible that snow cloud tidal debris could constitute a portion of the intermediate-velocity cloud (IVC) population that is thought to occupy the lower Galactic Halo \citep{leh22}.

In this paper we explore the process of snow-cloud tidal disruption by stars. As part of our exploration we simulate a suite of TDEs, using the ``affine'' formalism introduced by \cite{cart82,cart83,cart85}. While the affine description was initially developed to study the problem of stars being disrupted by black holes \cite[see also][]{khok93,fer06}, after a few straightforward adjustments we show that it is equally useful for describing snow clouds disrupted by stars. The main advantage of the affine approach is that computational demands for a given simulation are modest when compared to, say, smoothed-particle hydrodynamics \cite[e.g.][]{2012JCoPh.231..759P,lod20}; as a result we are able to build a library of simulations with a wide range of initial conditions, and thus study both individual and statistical properties of the collision products.

This paper is structured as follows. Section \ref{sec:hydrostatics} describes the snow-cloud structural model that we take as an initial condition for our affine calculations of TDEs. The basics of tidal disruption theory are described in Section \ref{sec:tidalbasic}, as is the affine model itself, together with some specifics relating to our numerical implementation. Models of TDEs typically assume that the disrupted body is initially weakly gravitationally bound to the disruptor, so that the initial orbit can be approximated by a parabola. We, on the other hand, consider unbound (hyperbolic) orbits and these populate a two-dimensional parameter space. Illustrative examples of individual events are shown in Section \ref{sec:repex}, where we evolve a small number of models of clouds swinging past a star with specified initial velocity and impact parameter. A library of simulations, covering a dense grid in the two-dimensional parameter space of initial orbits, is constructed and described in Section \ref{sec:library}. All of our simulations are undertaken with the stellar mass fixed at $M_\star=1\;{\rm M_\odot}$. The simulation library serves as input for statistical summaries, such as the kinematics of cloud debris streams, with results given in Section \ref{sec:stats}. Section \ref{sec:evolution} describes the likely evolution of a typical debris stream as it moves through the diffuse ISM, and in Section \ref{sec:connections} we consider possible connections to known astrophysical phenomena. Section \ref{sec:discussion} discusses some key issues that remain outstanding, and we wrap up with summary and conclusions in Section \ref{sec:conclusions}.

\section{Cloud hydrostatics}
\label{sec:hydrostatics}
In this first investigation of the tidal disruption of snow clouds, our main goal is to illustrate the general characteristics of the collision products for a single, simplified cloud model. That goal requires us to construct a large number of simulations covering a range of star-cloud impact parameters and relative velocities, so computation speed is important. Consequently (for reasons explained in \S\ref{sec:affine}) we adopt a single polytropic equation of state (EOS) with $P = K \rho^{\gamma}$, for fluid pressure $P$ and density $\rho$, given constants $K$ and $\gamma=1+1/n$, with $n$ being the polytropic index. The structural models of snow clouds presented by \cite{ww19} are also polytropic in the central regions of the cloud, where the partial pressure of \htwo\ is sub-saturated, and in those models it is the core that carries most of the mass. We therefore follow \cite{ww19} and employ an EOS that describes adiabatic convection in an effectively-monatomic, ideal gas --- namely a polytrope with index $n=3/2$. But in our model that EOS applies throughout the whole cloud. 

Although a wide range of snow cloud parameters may yield physically acceptable models, for the purposes of the present paper we seek a single, representative model. We therefore impose the following macroscopic requirements:
\begin{itemize}
\setlength\itemsep{1em}
\item[(i)]{The selected cloud mass is $\Mc = 3 \times 10^{-5}\, {\rm M_\odot}$. This value is the centre of the preferred mass range identified by \cite{2022MNRAS.513.2491T} from consideration of the lensing characteristics of a cosmological population of dense gas clouds.}
\item[(ii)]{As discussed in the Introduction, the mean cloud column-density is fixed at $\langle \Sigma \rangle \simeq 140\;{\rm g\,cm^{-2}}$, as determined by \cite{1999MNRAS.308..551W} from fitting to the observed properties of star-forming galaxies. Thus, for the mass specified in (i) we require a cloud radius $\Rc \simeq 0.78\;{\rm AU}$.}
\end{itemize}

\begin{figure}
\begin{center}
\includegraphics[width=0.45\textwidth]{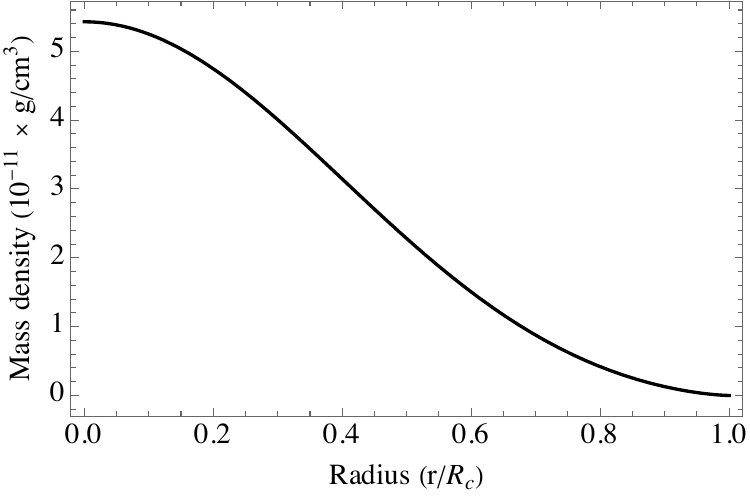}
\includegraphics[width=0.45\textwidth]{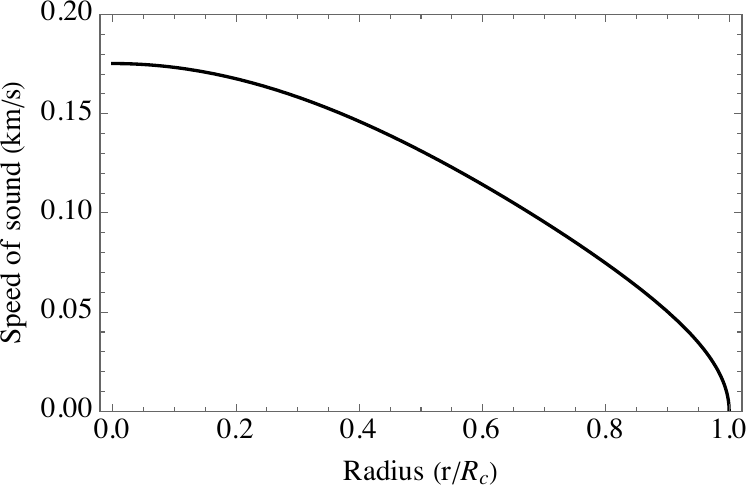}
\end{center}
\caption{Initial density profile, $\rho(r)$ (upper panel), and speed-of-sound, $c_s(r)$ (lower panel), as functions of radius for the hydrostatic model adopted in this work: a polytrope of index $n=3/2$, with mass $M_c=3\times 10^{-5}\,{\rm M}_\odot$ and radius $R_c = 0.777\;{\rm AU}$.}
\label{fig:massdens}
\end{figure}

For a polytrope with $n=3/2$ to yield these values of mass and radius we require \citep[see, for example,][]{ww19} a central density of $\approx 5.4\times10^{-11}\,{\rm g\,cm^{-3}}$ and a central pressure of $\approx 9.9\times10^{-3}\,{\rm dyne\,cm^{-2}}$. Again following \cite{ww19}, we assume that the composition, by mass, is 75\%\ \htwo\ and 25\%\ helium (no metals), so these values of central pressure and density correspond to a central temperature of approximately $5.1\;{\rm K}$. That is substantially lower than the central temperatures of the models presented by \cite{ww19} -- which range from $23\;{\rm K}$ to $100\;{\rm K}$ -- and therefore the fluid in our models has a central sound speed ($c_{s} \simeq 0.18\;{\rm km\,s^{-1}}$) that is smaller by a factor of 2 to 5. Readers should bear in mind this difference when, later, we consider the expansion of the debris stream, because the sound speed sets the free-expansion rate of the fluid. Figure \ref{fig:massdens} shows the density and sound speed of the fluid, for our model, as a function of radius within the cloud.

\section{Tidal disruption theory}
\label{sec:tidalbasic}
The strength of a tidal encounter with periastron distance $R_p$ is often quantified via the dimensionless parameter $\eta$, defined through
\begin{equation}\label{eq:disrad}
\eta^2 = \frac {\Mc} {M_{\star}} \left( \frac {R_p} {\Rc} \right)^{\!3} \approx \left( \frac {\Mc} {3 \times 10^{-5} M_{\odot}} \right) \!\!\left( \frac{R_p} {32 \Rc} \right)^{\!3}\!\! \left( \frac {M_{\odot}} {\Ms} \right)
\end{equation}
\citep[e.g.][]{rees88, koch94}. Disruption occurs if $\eta$ is less than some critical value that depends on the hydrostatic model of the disruptee. For an $n = 3/2$ polytrope, for instance, \cite{cart83} find $\eta_{\rm crit} \approx 1.2$ for parabolic orbits. This implies that, for a solar-mass star interacting with the model cloud described in the previous section, impact parameters $b \lesssim 28\;{\rm AU}$ should lead to disruption. As we will see, that is broadly in agreement with our numerical investigations (\S\ref{sec:stability}), which are primarily concerned with hyperbolic (unbound) orbits but which extend down to the parabolic limit. For hyperbolic orbits the cloud is moving faster than for a parabolic orbit with the same periastron, so there is less time for the tidal forces to effect distortions and a closer approach is required in order to disrupt the cloud.

\subsection{Affine cloud model}
\label{sec:affine}
In general, fully-fledged hydrodynamics codes are needed to describe the complex flows that arise in TDEs, such as shocks and accretion disk formation (see Table 1 in \citealt{lod20} for a survey on numerical approaches to TDEs). But in this paper we do not wish to examine the detailed properties of the fluid flows but, rather, to form statistical summaries of various aspects of the debris streams. Moreover, there is no fall-back and accretion in our circumstance, because the orbits are hyperbolic. Thus for our purposes a simpler treatment suffices, and we therefore adopt the semi-analytic affine model of \cite{cart82,cart83,cart85}. 

The affine approach builds on the incompressible models of \cite{c63}, tracking ellipsoidal contours of constant density that are allowed to undergo three-dimensional deformations under a time-dependent matrix. In general, one may consider an infinitesimal fluid parcel of the cloud located at coordinates $x_{i}$, in a Cartesian system oriented about the centre-of-mass (COM) of the \emph{star}. We introduce Lagrangian displacements $r_{i}(\boldsymbol{x},t)$, measuring the position of the fluid parcel relative to the COM of the cloud, through
\begin{equation}
x_{i} = X_{i} + r_{i},
\end{equation}
where ${\bf X}={\bf X}(t)$ is the position of the COM of the cloud. The key assumption of the affine model is that fluid parcels are not permitted to mix within the cloud, though the time dynamics are maintained at the fully non-linear level. More precisely, $\boldsymbol{r}$ is assumed to be linearly, though dynamically, related to the position of the same fluid parcel prior to deformation of the cloud (represented by tildes) through 
\begin{equation} \label{eq:transf}
r_{i} = q_{i j}(t)\; \tr_{j},
\end{equation}
where we employ the Einstein convention of summing over repeated indices. The nine components of $\bq$ are constrained by the Euler and continuity equations, and by the EOS of the fluid that makes up the cloud \citep{cart83}. We neglect reflex motion of the star, because $M_{\star} \gg M_{c}$, so that we do not need to recentre the coordinate system at each time step during numerical implementation.

The size and shape of the cloud at any time are fully described by the lengths of the three principal axes of the ellipsoid, which we designate $l_1$, $l_2$, $l_3$, with $l_1\ge l_2\ge l_3$; these are denoted (though without hierarchy) by $a_{i}$ in \cite{cart83}. Relative to the initial cloud radius, their values can be obtained from the ``configuration matrix'', ${\bf S}$, defined by $S_{i j} \equiv q_{i k} q_{j k}$: the squares of the lengths of the principal axes are given by the eigenvalues of ${\bf S}$ \citep{cart86}. To fully describe the shape of our deformed model cloud requires information on two axis ratios, but as far as shape goes we are mainly interested in characterising the general appearance of the debris to a distant observer: is it elongated (``cigar shaped'') or flattened (``disk-like'')? To differentiate between these two types of debris we make use of the ratio of the longest axis to the root-mean-square of the shorter axes: $M_{rms}\equiv l_1 \sqrt{2/(l_2^2+l_3^2)}$. Under this definition: $M_{rms}\sim 1$ for roughly spherical bodies ($l_1\sim l_2\sim l_3$); $M_{rms}\sim \sqrt{2}$ for flattened, disk-like structures ($l_1\sim l_2\gg l_3$); and, $M_{rms}\gg 1$ for highly elongated streams ($l_1\gg l_2 \ga l_3$).

Conservation of mass implies that the fluid density obeys a simple geometric relation \citep{cart86}:
\begin{equation} \label{eq:densityvar}
\rho(\boldsymbol{r},t) = \frac{\tilde{\rho}[\boldsymbol{q}^{-1}(t) \cdot \boldsymbol{r} ]}{ |\det{\boldsymbol{q}(t)|}}.
\end{equation}
The density distribution of the deformed cloud can therefore be anticipated immediately from its size and shape, because the radial profile simply follows that of the original hydrostatic equilibrium --- as shown in Fig.~\ref{fig:massdens}.

The COM trajectory and evolution equations for $\boldsymbol{q}(t)$ follow from the Euler-Lagrange equations, with the Lagrangian made up of contributions from the internal energetics ($L_{I}$), the ``external'' (COM) energetics ($L_{E}$), and the tidal coupling ($L_{C}$) \citep{cart85}:
\begin{equation} \label{eq:lag}
L = L_{I} +  L_{E} + L_{C}.
\end{equation}
The internal Lagrangian is given by $L_{I} = T_{I} - \Omega - {\cal U}$, with kinetic energy $T_{I}$, and potential terms related to self-gravity and to pressure. The kinetic energy can be expressed as $T_{I} = \tfrac{1}{2} \tilde{\mathcal{M}}\, \dot{q}_{ij} \dot{q}_{ij}$, where $\tilde{\mathcal{M}}$ is the scalar-quadrupole moment of the hydrostatic equilibrium state, evaluated from $\tilde{\mathcal{M}} \equiv \tfrac{1}{3} \int {\rm d\/}M \,\tilde{r}_{k} \tilde{r}_{k}$. The self-gravity term,  obtained from Poisson's equation via a Green's function, is $\Omega = -\tfrac{1}{2} G \iint {\rm d\/}M\,{\rm d\/} M' \left[ \left(r_{i} - r'_{i}\right) \left(r_{i} - r'_{i}\right) \right]^{-1/2}$, while ${\cal U}$ is the internal energy of the fluid:
\begin{equation} \label{eq:du}
{\rm d\/}\,{\cal U} =- \Pi \frac {{\rm d\/} \left(\det \boldsymbol{q} \right)} {\det \boldsymbol{q}} + \dot{Q}\, {\rm d\/}t.
\end{equation}
Here $\Pi = \int {\rm d\/}V P$ is the (perturbed) volume-integrated pressure, and $\dot{Q}$ (zero, in our case) describes the injection of heat \cite[e.g. from nuclear reactions;][]{cart82} or dissipation.  
The virial theorem ensures that $\tilde{\Omega} = - 3 \tilde{\Pi}$  \citep[e.g.][]{cart85} which can be used to fully eliminate pressure terms since
\begin{equation} \label{eq:pres}
\begin{aligned}
\Pi(t) &= \int {\rm d\/}V\; K \rho^{\gamma} \\
&= K \int ({\rm d\/} \tilde{V} \det \boldsymbol{q} )\, \left[ \tilde{\rho}^{\gamma} \left(  \det \boldsymbol{q} \right)^{-\gamma}  \right] \\
&= K \left( \det \boldsymbol{q} \right)^{1-\gamma} \int {\rm d\/} \tilde{V}\; \tilde{\rho}^{\gamma} \\
&= \tilde{\Pi} \left( \det \boldsymbol{q} \right)^{1-\gamma}
\end{aligned}
\end{equation}
for a polytrope (see equation 9 in \citealt{fer06}). Thus for a polytropic model we can evaluate $\Pi(t)$ directly from $\boldsymbol{q}(t)$, with no additional calculation, and it is primarily because of this simplification that we have employed a polytropic EOS.

The second term on the right-hand-side of \eqref{eq:lag} expands to $L_{E} = T_{E} - M_{c} \Phi_{\star}$, where $\Phi_{\star}= - G M_{\star} / \sqrt{X_{i} X_{i}}$ is the gravitational potential due to the star, and the kinetic energy is $T_{E} = \tfrac{1}{2}M_{c} \dot{X}_{i} \dot{X}_{i}$. These two terms account for the dynamics of the COM of the cloud in the gravitational field of the star. The third contribution to the total Lagrangian is the tidal coupling, $L_{C} = \tfrac{1}{2} \tilde{\mathcal{M}}\, C_{ij}\, q_{ik} q_{jk}$, where $\boldsymbol{C}$ is the Hessian of the stellar gravitational potential,
\begin{equation}
C_{ij} = - \frac{ \partial^2 \Phi_{\star}} {\partial X_{i} \partial X_{j}}.
\end{equation}

Introducing the gravitational self-energy tensor, $\Omega_{ij} = -q_{jk} \partial \Omega / \partial q_{ik}$, \cite{cart83} derive the overall equations of motion as\footnote{\cite{cart86} show how one may also include a viscous term in the above to properly handle shock fronts (essentially through a negative $\dot{Q}$). We ignore that complication here. Note also a typo in the indices in the second term on the right hand side of equation (3.17) in \protect\cite{cart83}.}
\begin{equation} \label{eq:eom}
\tilde{\mathcal{M}}\, \ddot{q}_{ij} = \tilde{\mathcal{M}}\, C_{ik}\, q_{kj} + \Pi \left(q^{-1} \right)_{ji} + \Omega_{ik} \left(q^{-1}\right)_{jk},
\end{equation}
and
\begin{equation} \label{eq:kinematics}
M_{c} \ddot{X}_{i} = - M_{c} \frac {\partial \Phi_{\star}} {\partial X_{i}} + \frac{1}{2} \tilde{\mathcal{M}}\, q_{\ell j} q_{k j} \frac {\partial C_{\ell k}} {\partial X_{i}}.
\end{equation}
The three terms appearing in \eqref{eq:eom} represent nondissipative contributions due to the tidal field, internal pressure, and  self-gravity of the cloud, respectively. The system is closed by specifying an EOS for the fluid --- polytropic, in our case, see \S\ref{sec:hydrostatics}. Equation \eqref{eq:kinematics} describes the acceleration of the COM of the cloud in the gravitational field of the star, accounting for the non-zero size of the cloud. Except near periastron, the second term on the right-hand side of equation \eqref{eq:kinematics} is small.

In its current form equation \eqref{eq:eom} is rather unwieldy, mostly because of the self-gravity term which does not, in general, admit a closed-form solution. The Green's function can, however, be evaluated in terms of simpler elliptic integrals for affine clouds \cite[e.g.][]{c87}. Using this fact together with \eqref{eq:pres}, equation \eqref{eq:eom} reduces to
\begin{equation} \label{eq:polyeom}
\begin{aligned}
\ddot{q}_{ij} &= C_{ik} q_{kj} + \left(\gamma-1\right)\tilde{\Psi} \left(\det{\boldsymbol{q}}\right)^{1-\gamma} \left(q^{-1}\right)_{ji} & \\
&- \frac{3}{2} \left(\gamma-1\right) \tilde{\Psi} \int^{\infty}_{0} {\rm d\/} \mu \frac{(\boldsymbol{S}+\mu \boldsymbol{I})_{k i}^{-1}} {\sqrt{\det\left( \boldsymbol{S} + \mu \boldsymbol{I} \right)}} q_{kj},
\end{aligned}
\end{equation}
where $\Psi ={\cal U} / \mathcal{M}$ is constant in the absence of dissipation and heat injection, so we have set $\Psi = \tilde{\Psi}$ in equation \eqref{eq:polyeom}. The kernel of the integral over the dummy variable $\mu$ involves the configuration matrix $\bf S$, introduced earlier ($S_{ij} = q_{ik} q_{jk}$), and the identity $\bf I$. 

We have now fully specified the problem: one must simultaneously solve the non-linear system of 9 transformation equations \eqref{eq:polyeom} for the components $q_{ij}$ and the three equations \eqref{eq:kinematics} for $X_{i}$ simultaneously, for a given polytropic cloud with particular values of the initial position and velocity. In practice the number of equations \eqref{eq:polyeom} that need to be solved can be reduced from 9 to 5, without loss of generality, if we restrict attention to orbits in a fixed plane, as the dynamics perpendicular to the orbital plane decouple from the in-plane dynamics \citep{cart86}, at least for irrotational clouds.

If the orbit of the cloud was followed inward from radial infinity at $t=0$ then initial conditions $q_{ij}(0) = \delta_{ij}$ and $\dot{q}_{ij}(0) = 0$ would be appropriate; however, it is not necessary to follow the evolution numerically when the separation between cloud and star is very large. In that regime the tidal tensor is tiny, so the cloud is only slightly distorted away from a sphere, and the linearised equations admit an analytic quasi-static solution when the evolution is slow compared to the free-fall time of the cloud, $\tau_{f\!f}$ (defined in \S\ref{sec:stability}):
\begin{equation} \label{eq:linearised}
q_{ij}(t) \approx \delta_{ij} + \frac{5}{4}\frac{\tilde{\cal M}}{\tilde{\Pi}}C_{ij}(t)
\end{equation}
\citep{cart83}. Furthermore, in this regime the COM trajectory is well approximated by that of a point mass, so $X_{i}(t)$ is easily evaluated for any specified impact parameter and relative velocity, with $C_{ij}(t)$ and $q_{ij}(t)$ following. The linear approximation \eqref{eq:linearised} thus allows us to specify the initial conditions, $q_{ij}(0)$ and $\dot{q}_{ij}(0)$, for the fully non-linear evolution at any desired star-cloud separation for which the linearised solution remains a good approximation.

\subsection{The tidal impulse approximation}
\label{sec:tidalimpulse}
The star-cloud interactions under consideration in this paper cover the two-dimensional space of impact parameter, $b$, and initial relative velocity, $u_i$, with the restriction that orbits should be unbound (hyperbolic) because the cloud and star are assumed to (initially) have no physical association. This situation is quite different to the parabolic orbit assumption that is commonly employed in TDE studies, for which the family of distinct orbits is described by a single parameter (the periastron distance). As a guide to the likely outcomes in our, larger parameter space, it is useful to have a simplified analytic solution of the affine model, as follows.

Clouds in hyperbolic orbits move at high speeds in comparison with those in parabolic orbits (having the same periastron distance), so a natural simplification is to treat the tidal forces as impulses acting on a spherical cloud. In that case the second and third terms on the right-hand side of equation \eqref{eq:polyeom} cancel, because the structure is as per the hydrostatic equilibrium, and we can integrate over time to obtain
\begin{equation} \label{eq:impulseapproximation1}
\dot{q}_{ij} = \int\! C_{ij}\, {\rm d}t,
\end{equation}
having used $q_{ij} = \delta_{ij}$ (cf. equation \ref{eq:linearised}).

Restricting attention to orbits in the plane $z=0$ permits further simplification without loss of generality, and the tidal tensor for that case is given in \cite{cart86}; four of the six off-diagonal components of $\boldsymbol{C}$ are zero. If we now neglect the deflection of the cloud COM by the gravitational field of the star then the position is a linear function of time and the integral of $\boldsymbol{C}$ is easy to evaluate. The remaining two off-diagonal elements of $C_{ij}$ are anti-symmetric about periastron and thus integrate to zero, leaving only the diagonal components of $\dot{q}_{ij}$ to be determined. We orient the coordinate system so that the trajectory of the cloud is $x=u_i t$, $y=b$ (constant). Here, and subsequently, the subscript $i$ denotes an initial ($t=0$) value, and the subscript $f$ denotes a value at the end of the simulation. The three diagonal components of $q$ then evaluate to
\begin{equation} \label{eq:impulseapproximation2}
\dot{q}_{xx}=0,\qquad {\rm and,} \qquad \dot{q}_{yy}=2\frac{GM_\star}{b^2 u_i} = -\dot{q}_{zz}.
\end{equation}
In this simplified treatment, then, the result of the interaction is that the cloud is not deformed in the direction parallel to the velocity ($x$), but it receives deformation-kicks of equal magnitude in the two orthogonal directions --- tensile for the in-plane direction ($y$), and compressive for the out-of-plane direction ($z$).

The magnitude of the deformation-kicks determines the fate of the cloud, according to whether the rate of deformation (\ref{eq:impulseapproximation2}) is large or small compared to the reciprocal of the free-fall time, $\tau_{f\!f}$, for the hydrostatic configuration of the cloud in isolation (see \S\ref{sec:stability}). If $|\dot{q}_{zz}| \ll \tau_{f\!f}^{-1}$ then the deformation is highly subsonic and the cloud will simply oscillate in response, whereas if $|\dot{q}_{zz}| \gg \tau_{f\!f}^{-1}$ the deformation is highly supersonic and the cloud will be disrupted by the tidal impulse. The dividing line between these two regimes thus provides us with an approximate boundary, in the $(b,u_i)$ plane, between collisions that are disruptive and those that are not. A precise criterion, based on energetic considerations, is presented later (\S\ref{sec:disruptioncriterion}) and used to assess the results of our numerical calculations.

It is convenient to introduce $\varpi= u_i\, \sqrt{b/4GM_\star}$, which is the initial relative velocity of cloud and star expressed in units of the speed of a parabolic orbit at periastron distance $b$.  That allows us to express our disruption criterion in a compact form:
\begin{equation} \label{eq:hyperboliccriterion}
|\dot{q}_{zz}|\, \tau_{f\!f} \simeq \frac{1}{\eta\, \varpi} \ga 1,
\end{equation}
where we have made use of the parameter $\eta$ introduced in equation (\ref{eq:disrad}).

If a collision is disruptive, the evolution implied by our simplified model is as follows. The cloud flies past the star on the orbital timescale $t_{orb} = 2\,b/u_i$, and subsequently it deforms on the timescale $\sim \varpi^2\,t_{orb}$. Clearly the deformation timescale is longer than the orbital timescale for $\varpi>1$ (hyperbolic collisions), consistent with our application of the impulse approximation. The deformation itself involves a flattening of the out-of-plane ($z$) structure of the cloud, and a simultaneous stretching of one of the in-plane dimensions ($y$) at the same rate, while leaving the third dimension ($x$) unchanged. In other words: an oval pancake arises. Pancaking implies that the fluid pressure reaches high values compared to those in the original hydrostatic configuration. Thus, similar to the disruptions that have been described for parabolic orbits \citep[e.g.][]{cart83}, there will be a bounce, followed by an expansion in $z$ that is as rapid as the one already underway in $y$. Thus our tidal impulse approximation leads us to expect that, after a few deformation timescales, a disrupted cloud has the form of an ellipsoid with $l_1=l_y\sim l_2=l_z \gg l_3=l_x$.

Although the foregoing analysis is highly simplified we will see later (\S\ref{sec:library}) that it provides a useful, rough sketch of what is actually found in the full numerical calculations.

\subsection{Numerical implementation}
\label{sec:numerics}
The system \eqref{eq:polyeom} is stiff, owing mostly to the determinant terms which can have large gradients even for relatively small changes in the components $q_{ij}$. The size of the time-step (typically of order $\sim 10^{-4}$ in dimensionless units) is adaptively controlled by demanding that the local error, in dimensionless units, be at most one part in $10^{10}$ --- ensured using local routines within \textsc{Mathematica.}\footnote{https://www.wolfram.com/mathematica} It is convenient to choose the units of mass, length and time in our simulation as $10^{-3}\,\text{M}_{\odot}$, $0.1\;\text{AU}$ and  and $1\;\text{yr}$, respectively \cite[see also Appendix B in][]{iv01}. The chosen mass scale is approximately the geometric mean of the star and the cloud, so as to minimise the spread of dimensionless numbers used within the simulation, while the length scale is chosen so as not to exceed the static cloud radius.

Finally, we note that the integral within equation \eqref{eq:polyeom} can be expressed in terms of incomplete elliptic integrals of the first and second kind using formulae given in Appendix A of \cite{cart86}; this allows for a faster evaluation as there are efficient, well-calibrated methods for calculating these latter integrals \cite[see, e.g.,][]{carl95}.

\subsection{Stability checks}
\label{sec:stability}
As a check on the stability of our numerical implementation, we undertook some test calculations to verify that the scheme behaves as expected. Recall that the free-fall time, $\tau_{f\!f}$, is the timescale on which a pressureless, self-gravitating fluid will collapse to a point when released from rest; we have \cite[e.g.][]{bintrem}
\begin{equation} \label{eq:freefall}
\tau_{f\!f} = \frac{1}{4} \sqrt{\frac{3 \pi} {2 G \bar{\rho}}} \approx 22 \;\text{yr},
\end{equation}
for volume-averaged density $\bar{\rho}$. In the case where the tidal field of the star is weak, or absent, our model clouds should not exhibit any large changes in their structure even over timescales that are long compared to $\tau_{f\!f}$.
 
\begin{figure}
\begin{center}
\includegraphics[width=0.45\textwidth]{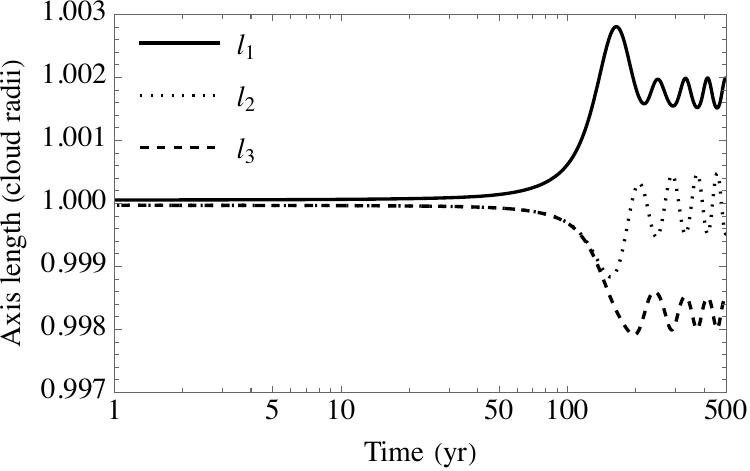}
\end{center}
\caption{Evolution of the three principal axes of the cloud (see legends) for a non-disruptive collision with a very large impact parameter ($b=200\;{\rm AU}$). The initial horizontal separation is $600\;{\rm AU}$, and the initial velocity is $u_i=20\;{\rm km\,s^{-1}}$ in that direction.}
\label{fig:hydrostab}
\end{figure}

Figure \ref{fig:hydrostab} shows the evolution of the principal axes of the cloud for a cloud-star interaction with an initial, relative speed of $20\;{\rm km\,s^{-1}}$ and an impact parameter of approximately $200\;{\rm AU}$; this interaction is well into the regime where no disruption should occur. The initial separation between cloud and star is $600\;{\rm AU}$ (in $x$), which is covered in approximately $150\;{\rm yr}$ ($\sim 7 \tau_{f\!f}$).

\begin{figure}
\begin{center}
\includegraphics[width=0.45\textwidth]{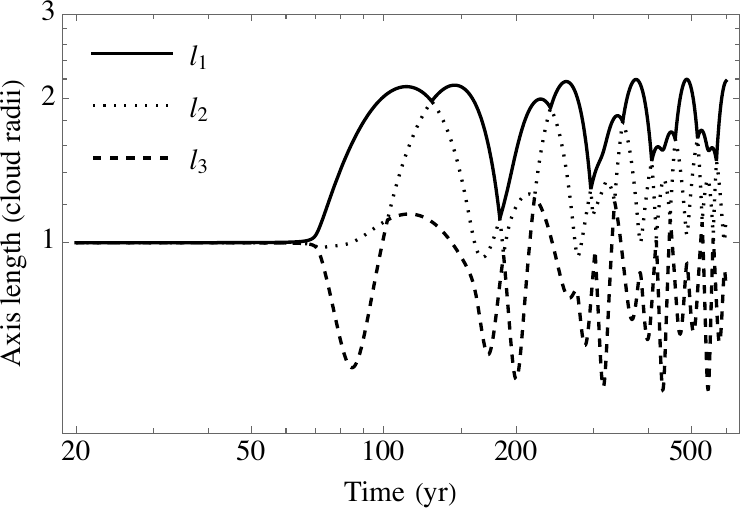}
\end{center}
\caption{Similar to Fig.~\ref{fig:hydrostab}, but with impact parameter $b=14\;{\rm AU}$ and initial velocity $u_i=40\;{\rm km \,s^{-1}}$. This combination does not yield a disruption, but the tidal interaction is strong and large-amplitude oscillations arise, post periastron, in all three axes.}
\label{fig:hydrobord}
\end{figure}

As expected, the principal axes of the cloud change little until it is close to periastron, indicating that at early times the original hydrostatic structure is maintained to a good approximation --- there is just a slight, gradually increasing deformation, as per the linear approximation given in equation (\ref{eq:linearised}). The tidal forces reach maximum at periastron, and it is around then that the cloud exhibits its largest deformation. After moving away from the star, though, the cloud is left with a permanent deformation, being almost $0.2$\%\ smaller in the out-of-plane direction, and commensurately larger with respect to one of the in-plane axes.
Post periastron the cloud also exhibits oscillations, the small-amplitude ($\sim 0.01$\%) of which is set by the maximum magnitude of the tidal tensor, in all three axis lengths; these persist for the duration of the simulation, as expected in the absence of damping. In any real cloud such oscillations would be damped by the viscosity of the fluid.

Despite the lack of viscosity in the simulation, Figure \ref{fig:hydrostab} demonstrates that any numerical instabilities in the code -- which are likely to be present at some level, due to roundoff error etc. -- have not grown to a noticeable level (much less than $0.1$\%), even over a time corresponding to over thirty free-fall times. For our purposes this level of performance is more than sufficient, because tidal disruptions (which all have much smaller impact parameters than this event) do not require the evolution to be followed for so long.

\begin{figure*}
\includegraphics[width=\textwidth]{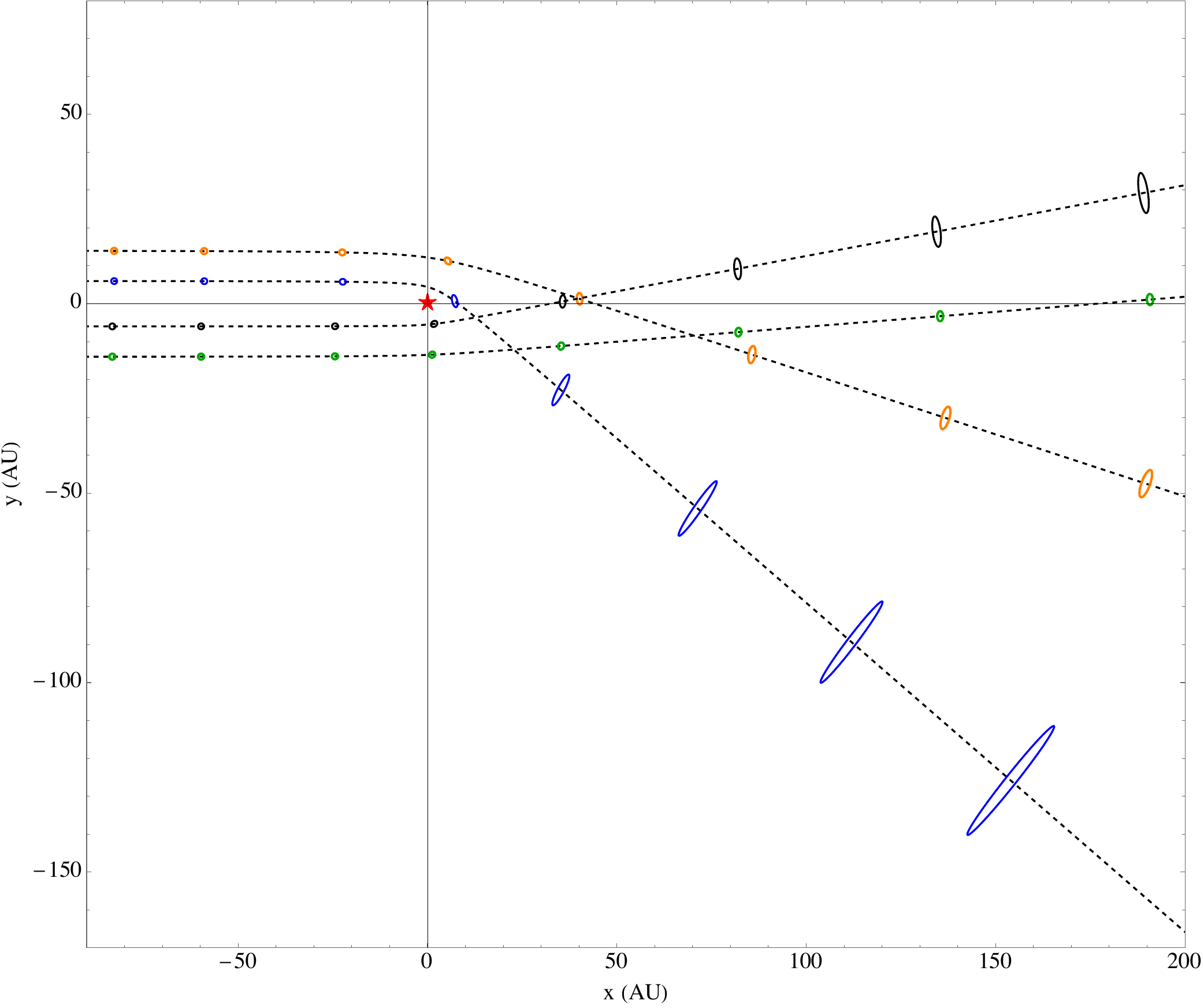}
\caption{Four examples of star-cloud tidal interactions, as computed with the affine model. Cloud centre-of-mass trajectories are shown with black, dashed curves, and the coloured lines show the ($z=0$ slices of the) cloud surfaces for eight well-separated snapshots in time. In all cases the orbital plane is $z=0$ with the star fixed at the origin. The initial conditions for the four examples shown are ${\bf X}(0)=(x_i, y_i, 0)$, with $x_i=-200\;{\rm AU}$ and $y_i=\pm b$, while $\dot{{\bf X}}(0)=(u_i, 0, 0)$ and (top to bottom, on the left of the figure): $b=14\;{\rm AU}$, $u_i=20\;{\rm km\,s^{-1}}$ (orange); $b=6\;{\rm AU}$, $u_i=20\;{\rm km\,s^{-1}}$ (blue); $b=6\;{\rm AU}$, $u_i=40\;{\rm km\,s^{-1}}$ (black); and, $b=14\;{\rm AU}$, $u_i=40\;{\rm km\,s^{-1}}$ (green).
}
\label{fig:diagramT}
\end{figure*}

By contrast with the interaction described above, Figure \ref{fig:hydrobord} shows a case where the impact parameter/initial velocity combination ($b=14\;{\rm AU}$, $u_i=40\;{\rm km\,s^{-1}}$) place the cloud close to (but outside) the disruption boundary (see \S\ref{sec:disruptioncriterion}). In this case we again see little evolution in the principal axis lengths until closest approach (occuring at $t\simeq 75\;{\rm yr}$). However, with a much closer approach to the star in this case, the response of the cloud is strong and the axis lengths exhibit large amplitude oscillations. Further decreasing the impact parameter and/or the relative speed of the interaction can give rise to disruptions, which exhibit secular growth in the axis lengths --- as we will see in the next section.

\section{Examples of disruptions}
\label{sec:repex}
In this section we present results that illustrate cloud disruptions simulated with the affine model. Figure \ref{fig:diagramT} shows trajectories and $z$-plane slices of the  cloud surfaces for three such disruptions, along with the non-disruptive encounter shown in Fig.~\ref{fig:hydrobord} for comparison. The four examples shown all constitute interactions with orbits in the $z=0$ plane, with initial velocities in the $+x$ direction, and thus the initial $y$ values are simply $y_i=\pm b$ (the impact parameter), with either $b=6\;{\rm AU}$ or $b=14\;{\rm AU}$. The two different choices of impact parameter combine with two different choices of initial velocity, $u_i=20\;{\rm km\,s^{-1}}$ or $u_i=40\;{\rm km\,s^{-1}}$, to yield the four cases shown. For clarity we have placed both of the $u_i=20\;{\rm km\,s^{-1}}$ trajectories at $y_i=+b$, and both of the $u_i=40\;{\rm km\,s^{-1}}$ trajectories at $y_i=-b$. For all these calculations the initial value of the $x$ coordinate is far outside the domain of the plot, at $x_i=-200\;{\rm AU}$. 

Figure \ref{fig:diagramT} shows that events with smaller impact parameters lead to stronger disruptions, as expected because the tidal tensor increases rapidly as the cloud gets closer to the star --- as the inverse-cube of the separation. The figure also demonstrates that faster flybys lead to weaker disruptions. That too is unsurprising: if the initial relative velocity is doubled then the cloud experiences the gravitational tides of the star for roughly half as long, resulting in correspondingly smaller deformations. 

\subsection{Disruption criterion}
\label{sec:disruptioncriterion}
A perturbed cloud generally oscillates in a superposition of normal modes \citep{pt77}, which cause the volume to fluctuate as the simulation progresses. Here we are interested in disruptions, for which the volume is expected to grow without bound; but in the presence of oscillations that is difficult to ascertain in borderline cases --- cf. Fig. \ref{fig:hydrobord}, which would have been less clear-cut had the simulation only run for $200\;{\rm yr}$. A numerically tractable definition is challenging to formulate, because oscillations are not volume-preserving, and the form in which they are manifest changes from simulation to simulation.

We have experimented with different criteria to decide on whether a cloud is disrupted or not. For example: we could say that a disruption occurs if the derivative of the determinant of the deformation matrix $\boldsymbol{q}$ is positive over the last free-fall time of the simulation. The free-fall time is about a quarter period of the dominant excited modes --- see Figs.~\ref{fig:hydrostab} and \ref{fig:hydrobord}. Such a criterion attempts to capture only cases which display a genuine, secular growth in volume. While this definition inevitably miscategorises some borderline cases (e.g. mildly disrupted clouds in a contracting phase at the end of the simulation), it proved effective in most instances.

However, a physically robust criterion is one based on energetics, as follows. In terms of the quantities introduced in \S\ref{sec:tidalbasic}, and the total energy internal to the cloud (i.e. measured in its COM frame),
\begin{equation}
{\cal E} = \tfrac{1}{2} \tilde{\mathcal{M}}\, \dot{q}_{kj} \dot{q}_{kj} + \Omega + {\cal U},
\label{eq:totalenergy}
\end{equation}
the full Hamiltonian of the system is 
\begin{equation}
{\cal H} = \tfrac{1}{2}M_{c} \dot{X}_{j} \dot{X}_{j} + M_{c} \Phi_{\star} + {\cal E} - \tfrac{1}{2} \tilde{\mathcal{M}} C_{\ell j} q_{\ell k} q_{jk},
\label{eq:hamiltonian}
\end{equation}
which is strictly conserved ($\dot{\cal H} = 0)$ in the absence of dissipation (e.g. viscosity) or heat injection (e.g. from chemical reactions).  To decide whether a cloud has been disrupted we focus attention on the behaviour at late times, when the cloud is far from the star and the tidal coupling term in equation \eqref{eq:hamiltonian} becomes negligible. In that limit the Hamiltonian is a sum of two parts that are separately conserved: one part relating to the motion of the COM (the first two terms on the right-hand-side of equation \ref{eq:hamiltonian}), which then evolves in just the same way as a point-like particle; and ${\cal E}$ which describes the internal properties of the cloud, wherein the star no longer plays a role. Thus, although the potential energy, kinetic energy, and pressure terms when taken individually are each oscillatory, at late times their sum, ${\cal E}$, is constant to a good approximation:  ${\cal E}\simeq {\cal E}_f$. We require that ${\cal E}_f>0$, at the end of the simulation, for a cloud to qualify as ``disrupted''.

In any real cloud the oscillations would, of course, dissipate as a result of the viscosity of the fluid, and the late-time configuration of a cloud that is \emph{not} disrupted would again be a hydrostatic equilibrium with $\Omega=-3\Pi$, but with a radius larger than the original cloud by a factor $\tilde{\Omega}/\Omega$ (though this will not, in general, be the same $\Omega$ achieved in a simulation without viscosity).

In combination with the solution (\ref{eq:impulseapproximation2}), derived using the impulse approximation (\S\ref{sec:tidalimpulse}), the criterion ${\cal E}_f>0$ yields a more accurate description of the expected boundary in the $(b,u_i)$ plane between disrupted and non-disrupted clouds:
\begin{equation}
b^2\,u_i = 2\, \frac{GM_\star}{\sqrt{\tilde{\Psi}}}.
\label{eq:disruptionboundary}
\end{equation}
From our hydrostatic model of the cloud we can evaluate $\tilde{\Psi}$; the result is $\tilde{\Psi}=1.056\times 10^{-17}\;{\rm s^{-2}}$, which yields the numerical relation $b{\rm (AU)}\simeq 60.4/\sqrt{u_i{\rm (km\,s^{-1}})}$ in the case of a solar-mass star. The functional form of this result is of course identical with the form implied by equation (\ref{eq:hyperboliccriterion}), and the two expressions differ only in the constant of proportionality. 

We emphasise that the result given in (\ref{eq:disruptionboundary}) applies specifically to the tidal impulse solution (\ref{eq:impulseapproximation2}), and is therefore only approximate. By contrast the condition ${\cal E}_f>0$, which is the criterion used in our numerical work, involves no approximation. Nevertheless, in \S\ref{sec:library} we will see that the approximate boundary (\ref{eq:disruptionboundary}) is quite accurate under circumstances where the impulse approximation can sensibly be applied.

\subsection{Drift velocities}
\label{sec:driftvel}
The rate at which a disrupted cloud expands can be gauged by the \emph{drift velocities} of surface elements relative to the centre, in the following way. First we find the velocity of a general fluid element, relative to the cloud COM, by differentiating equation \eqref{eq:transf} with respect to time:
\begin{equation} \label{eq:drift}
\delta v_{j} = \frac{{\rm d} q_{jk}} {{\rm d}t}\tilde{r}_{k}.
\end{equation}
Then we locate the maximum value over all the surface elements of the cloud, and thus the drift speed.

\begin{figure}
\begin{center}
\includegraphics[width=0.48\textwidth]{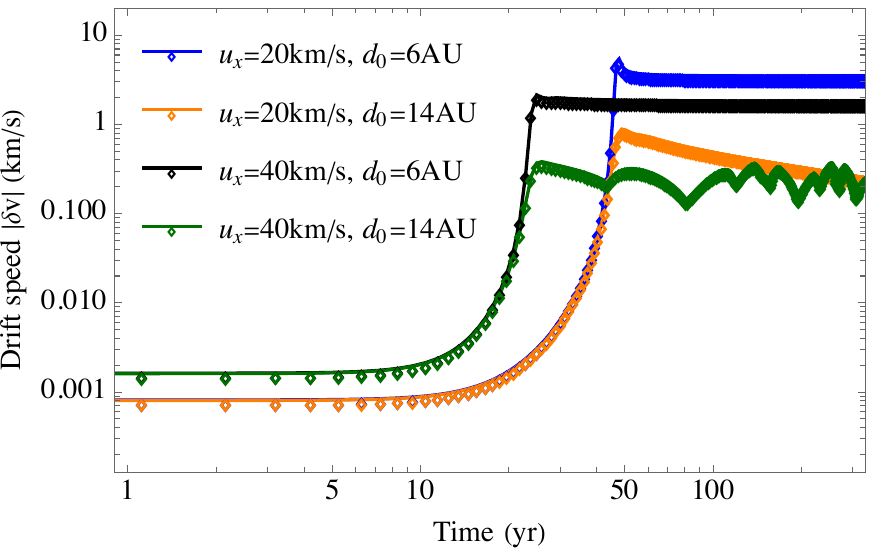}
\end{center}
\vskip -0.6cm
\caption{Drift speeds, $|\delta \boldsymbol{v}|$, as a function of time for the simulations shown in Fig.~\ref{fig:diagramT} (see plot legends). Plot markers indicate numerically sampled points in time, with the curve showing the linear interpolation.}
\label{fig:driftsample}
\end{figure}

Figure \ref{fig:driftsample} shows the evolution of the drift speeds $|\delta \boldsymbol{v}|$ through the four simulations presented earlier in this section (Fig.~\ref{fig:diagramT}), as determined from $22^2$ uniformly distributed points over the cloud surface (visually indistinguishable results are obtained if $20^2$ or $24^2$ points are used instead). Note that the drifts are non-zero even at the start of the simulation because of the initial conditions \eqref{eq:linearised}. We see that the largest drift speed is achieved in the case that combines the lowest relative velocity of cloud and star ($20\;{\rm km\,s^{-1}}$) with the smallest impact parameter ($6\;{\rm AU}$) (Fig.~\ref{fig:diagramT}), viz. $|\delta \boldsymbol{v}| \approx 3.5\;{\rm km\,s^{-1}}$ at late times (blue curve). The disruption is clearly hypersonic, at $\simeq 20\times c_{s}(0)$. Increasing the impact parameter to $14\;{\rm AU}$ (orange curve) leads to a case which is right at the border of disruption and illustrates the competing nature of expansion and oscillation. The cloud reaches a supersonic drift velocity ($|\delta \boldsymbol{v}| \approx 0.8\;{\rm km\,s^{-1}}$), though this decreases with time as the oscillations enter a contracting phase in opposition to the secular expansion. Since the mean density has significantly decreased by this stage however, the oscillation period is longer (having increased by a factor $\sim \rho^{-1/2}$) than the simulation timescale. Had we extended the domain by a factor of a few, a cycle of local minima in $\delta \boldsymbol{v}$ would be observed, dipping to $\sim$sonic values, followed by rises to supersonic values. Viscosity, absent from our simulations, would be especially important in such cases.

For the two cases that start with a higher initial relative velocity ($u_i=40\;{\rm km\,s^{-1}}$) between star and cloud (black and green curves in Fig.~\ref{fig:driftsample}), the drift speed exhibits the same form at early times ($|\delta \boldsymbol{v}| \propto t^2$), with this quadratic growth halting earlier because periastron is reached in approximately half the time. Moreover, because the star's tides have less time to act on the cloud, the case that disrupts ($b=6\;{\rm AU}$, black curve) attains a final drift speed that is lower by a factor $\sim 2$ relative to the corresponding case with $u_i=20\;{\rm km\,s^{-1}}$. The relative magnitudes of the drift velocities in these three cases of disruption can be seen to be consistent with the relative sizes of those clouds in the final snaphots in Figure \ref{fig:diagramT}. 

In the one instance of a non-disruptive interaction (green curve; $b=14\;{\rm AU}$, $u_i=40\;{\rm km\,s^{-1}}$) the drift speed exhibits large-amplitude variations at late times --- reflecting the complex oscillatory behaviour of the cloud axes seen in Fig.~\ref{fig:hydrobord}, with energy flowing back and forth amongst the three internal forms (pressure, kinetic energy, and self gravity). In the following sections of this paper, where we are interested in the late-time secular expansion of disrupting clouds, we employ a time-averaged (over the last two free-fall times) measure of the drift speed.

\section{A collision library}
\label{sec:library}
In the preceding section we showed a small number of examples of tidal disruptions arising from star-cloud interactions. But the main aim of this paper is to arrive at a statistical description of the debris streams, and an essential part of that endeavour is the construction of a library of simulations that provides a clear picture of the properties of the tidal debris as a function of the relative velocity and impact parameter of the collision. That characterisation is given in this section. 

All the quantities we show here are evaluated on a uniform, 2-dimensional grid of points spanned by the impact parameter, $b$ [$2 \le b\,{\rm (AU)} \le 35$, with a spacing of $1/3\;{\rm AU}$], and initial relative speed, $u_{i}$ [$4 \le u_{i}\,{\rm (km\;s^{-1})} \le 220$, with a spacing of $2\;{\rm km\,s^{-1}}$] --- a total of almost 11{,}000 points. Low velocity collisions have been excluded from the grid because they correspond to orbits that are initially bound to the star, whereas in this work we are addressing the case where star and cloud initially have no physical association. We have also excluded orbits with such small impact parameters that a physical collision would occur between star and cloud; such a collision could not be described by the formalism used in this work. In both cases the exclusions are rendered simplistically, with a fixed velocity cutoff and a fixed impact parameter cutoff, because these regions of parameter space contribute very little to the overall statistics. We will see later that our grid does actually still include (at the low velocity end) some bound orbits, which we can exclude consistently by checking the eccentricity as a function of time.

Each simulation was initiated at a star-cloud separation of $-x_i=200\;{\rm AU}$, which is much larger than the largest impact parameter on our grid of models, and large enough to ensure that the linearised model (equation \ref{eq:linearised}) provides a good approximation for the initial state of the cloud. Post periastron, each simulation was followed out to a final cloud-star radial separation that is larger than the initial separation, so that the tidal tensor has shrunk to a tiny fraction of its value at periastron, and the final energy in the cloud COM frame, ${\cal E}_f$, can be accurately estimated from equation (\ref{eq:totalenergy}).

\begin{figure}
\begin{center}
\includegraphics[width=0.45\textwidth]{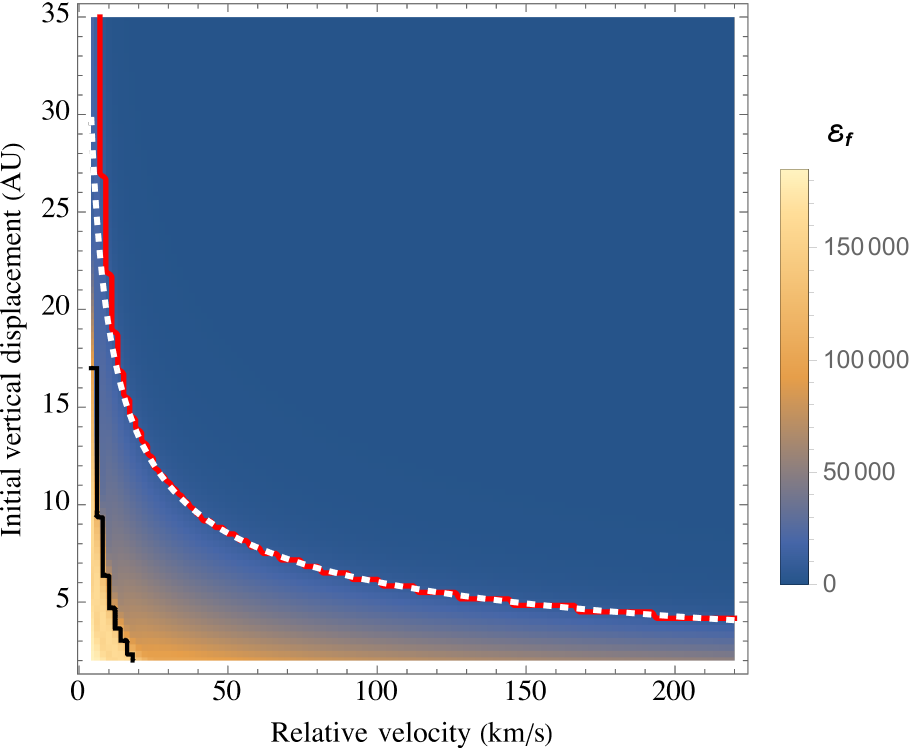}
\end{center}
\caption{The final total energy, ${\cal E}_f$, of the cloud as seen in its COM frame (equation \ref{eq:totalenergy}), as a function of the relative speed ($u_i$, horizontal axis), and impact parameter ($b$, vertical axis) of the collision. The final total energy is expressed in units such that the initial total energy is ${\cal E}_i=-1$. The boundary between clouds that survive the stellar fly-by (${\cal E}_f < 0$), in the upper-right of the diagram, and those that are disrupted, which lie closer to one of the axes and have ${\cal E}_f > 0$, is shown by the solid red line. The white dashed line shows our analytic approximation to that boundary, equation \eqref{eq:disruptionboundary}, obtained using the tidal impulse approximation. The solid, black line instead shows cases which, although initially hyperbolic ($e_{i}>1$), become bound post-interaction ($e_{f}<1$).}
\label{fig:totalenergy}
\end{figure}

Perhaps the most fundamental issue we need to address is which orbits are disruptive, so ${\cal E}_f$ is the first quantity that we take stock of. In units such that the initial total energy is ${\cal E}_i=-1$, strong disruptions correspond to ${\cal E}_f \gg 1$. The variation of ${\cal E}_f$ with relative speed, $u_i$, and impact parameter, $b$, is shown in Figure~\ref{fig:totalenergy}, where it can be seen that the strong disruptions are concentrated near the axes, reaching energies $\mathcal{E}_f \gtrsim 10^{5}$. That is easily understood, as follows. A small impact parameter guarantees that the cloud approaches very close to the star, hence the tidal interaction becomes very large and the cloud is strongly disrupted. Similarly, if the relative speed is sufficiently small then the cloud will plunge in towards the star, reaching a periastron distance that is much smaller than the impact parameter, and again a strong disruption will ensue.

The red line plotted on Figure~\ref{fig:totalenergy} shows the boundary between disruptive and non-disruptive collisions, as determined from our simulations: all clouds to the lower-left of that locus are disrupted, whereas all clouds to the upper-right survive. Also plotted (dashed white line) is our analytic estimate of the disruption boundary, derived in the impulse approximation (equation \ref{eq:disruptionboundary}). Although stemming from a rather crude model, the latter lies very close to our numerically determined boundary in most of the parameter space covered in Fig.~\ref{fig:totalenergy}, especially when the relative velocities of star and cloud are high. That indeed is the region where we expect the tides to be most impulsive, and thus the approximation to be most accurate. We can quantify that expectation, because our derivation in \S\ref{sec:tidalimpulse} assumed (i) that the collision is over before the cloud has time to evolve ($t_{orb} \ll \tau_{f\!f}$), and (ii) the COM trajectory is approximately linear (hyperbolic collision, with eccentricity$\,\gg 1$). Together these conditions exclude a thin region at the left of Figure~\ref{fig:totalenergy} from the domain of validity of our analytic derivation of the boundary, and we find that the dashed white curve should not be trusted for $u_i\la 16\;{\rm km\,s^{-1}}$ ($b\ga 15\;{\rm AU}$).

During a cloud-star encounter there is always some increase in energy, ${\cal E}_f > {\cal E}_i$, at the expense of energy in the orbit of the cloud COM around the star --- the eccentricity of that orbit decreases. If the initial eccentricity of the orbit is not too large that may lead to tidal capture of the cloud \citep[similar to, but easier than, one star being tidally captured by another;][]{1975MNRAS.172P..15F, pt77}. And, of course, tides act to circularise any bound orbits. Those are potentially interesting aspects of cloud-star interactions, but they are beyond the scope of the present work and in this paper we do not seek to study either tidal capture or tidal circularisation. However, with our simulations we can easily track the eccentricity of the orbit as a function of time. The solid black line in the bottom-left corner of Fig.~\ref{fig:totalenergy} shows the region where the eccentricity drops below unity; this region is excluded from subsequent analysis.

\begin{figure}
\begin{center}
\includegraphics[width=0.45\textwidth]{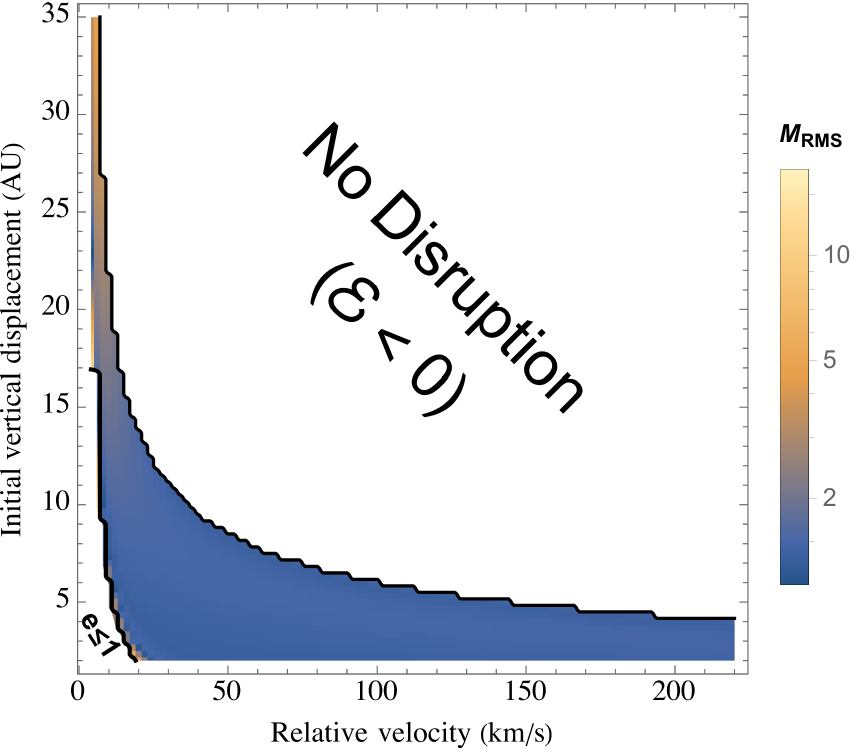}
\end{center}
\caption{As for Fig~\ref{fig:totalenergy}, but showing the final axis ratio of the cloud, $M_{rms}(u_i,b)$.}
\label{fig:machnums}
\end{figure}

Figure \ref{fig:machnums} shows how the final axis ratio of the cloud varies with $b$ and $u_i$. Recall that our axis ratio definition is one that averages over the 3D structure, with $M_{rms} = l_1 \sqrt{2/( l_2^2 + l_3^2)}$ (and $l_1\ge l_2\ge l_3$). For most of the velocity range, the debris forms a flattened (disk-like) or quasi-spherical structure, with $M_{rms}\la \sqrt{2}$, not an elongated stream. In particular, as we will see in the following section, that is true for relative velocities ($u_i\sim \sigma_{e\!f\!f}$) that are typical of star-cloud interactions, and therefore most of the debris streams should be of that type. Somewhat larger elongations can arise when the relative velocity is small -- e.g. $M_{rms}$ values of $\gtrsim 30$ at the upper left of the figure -- as the interaction approaches the parabolic limit, with orbital eccentricity $e=1$, that forms the boundary of the set of unbound orbits. We exclude bound orbits ($e<1$) from consideration. Had we included the $e\leq1$ region the overall statistics would not change much because such interactions are very rare --- e.g. direct calculation reveals that the disruption rate would be increased by only $\approx 0.2\%$ (see \S\ref{sec:colr}).

\begin{figure}
\begin{center}
\includegraphics[width=0.45\textwidth]{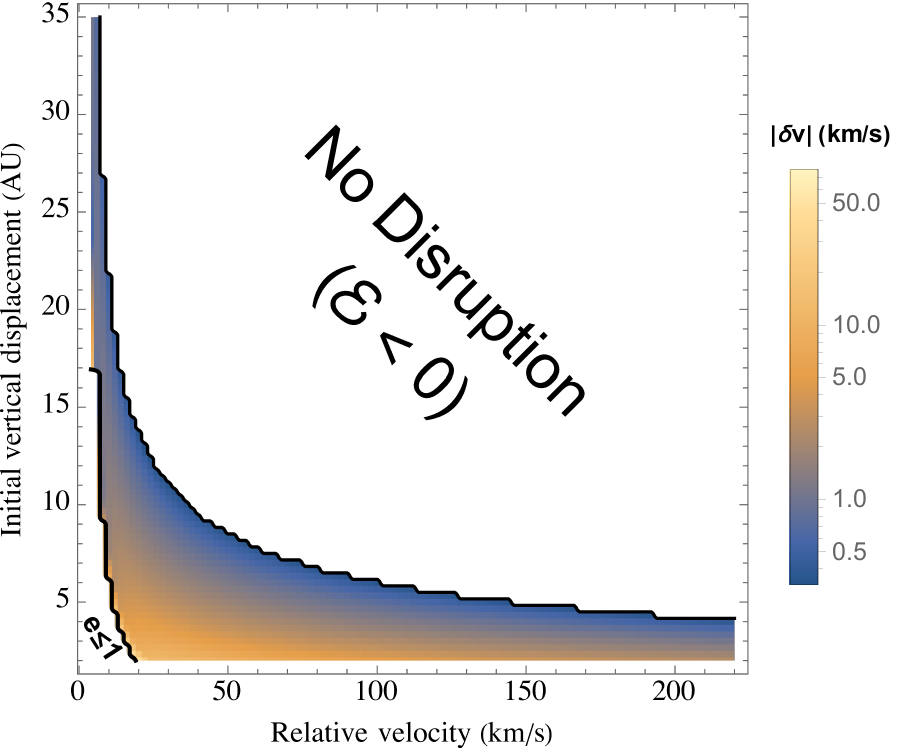}
\end{center}
\caption{As for Fig~\ref{fig:totalenergy}, but showing drift speeds, $|\delta \boldsymbol{v}(u_i,b)|$.}
\label{fig:driftvels}
\end{figure}

Similarly, the drift speeds of disrupted clouds are shown in Figure~\ref{fig:driftvels}, as a function of $b$ and $u_i$. The structure in that plot largely echoes what is seen in the energies (Figure \ref{fig:totalenergy}), with the fastest drifts (as much as $\sim 80\;{\rm km\,s^{-1}}$, near the origin) corresponding directly to the largest energies, and the slowest drifts achieved for borderline cases with small $\mathcal{E}_f > 0$.

\section{The statistical properties of debris}
\label{sec:stats}
Having considered the raw simulation outputs in the previous section, we now move on to the main goal for this work: deducing the statistical properties of a hypothetical population of disrupted clouds. That requires us to evaluate several multi-dimensional integrals, as we need to sum over contributions associated with the three velocity components of both cloud and star, the (two-dimensional) impact vector, and the orientation of the orbital plane. For most of the statistics of interest it is only the distribution of relative velocities (cloud relative to star, say) that affects the result, and for those cases the dimensionality of the numerical integrals can be reduced by recasting the problem into sum- and difference velocities. However, for evaluating the kinematic distributions of the debris streams (\S\ref{sec:kinematics}) no such simplification is possible. Throughout this section, multi-dimensional integrals are computed using a Monte Carlo method with $10^7$ points, implying errors below the level of $0.1$\%\ for global estimates and $1$\%\ on the values of $\sim10^2$ bins in a probability distribution.

\subsection{Initial kinematics of clouds and stars}
\label{sec:priors}
To explore the statistics of clouds having undergone tidal disruption by stars it is necessary to specify the initial kinematics of both populations. Although the stellar velocity distribution is known from observations, it is unclear what kinematics should apply to a population of cold, dense gas clouds, as there is currently no reliable theoretical model for the origin of this dark matter component. Bearing in mind also the other uncertainties and approximations inherent in the modelling (e.g. \S\ref{sec:hydrostatics}), we adopt the same distribution function that we use for the stars in this initial exploration of disrupted clouds.

\begin{figure}
\begin{center}
\includegraphics[width=0.45\textwidth]{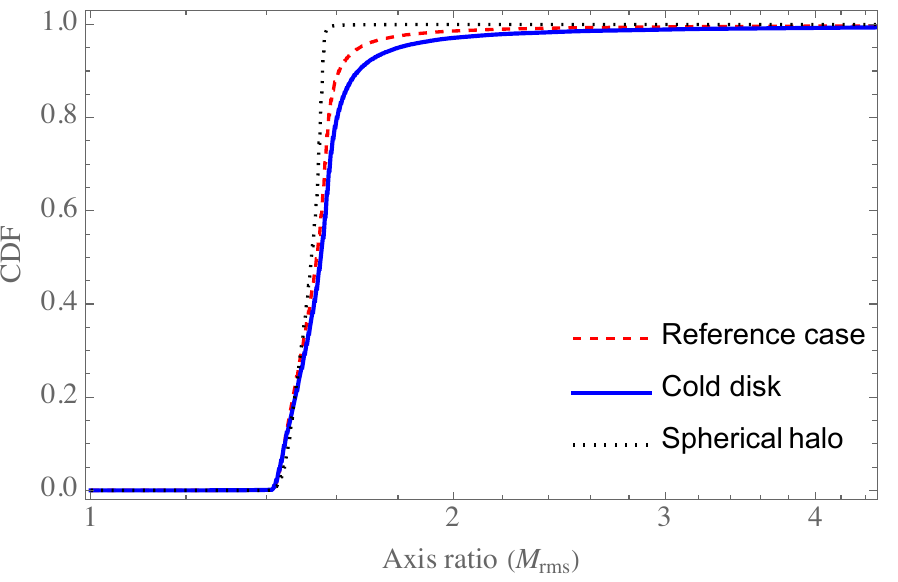}
\end{center}
\caption{The cumulative distribution function for the axis ratio, $M_{rms}$, of the debris stream for our reference case (red, dashed line), where the cloud velocities are distributed in the same way as the population of thin-disk stars. Also shown for comparison are results appropriate to clouds in a non-rotating, spherical distribution (black, dotted line) and clouds in cold-disk rotation (blue line).}
\label{fig:machcdf}
\end{figure}

We are interested in describing the solar neighbourhood, for which a suitable model is a triaxial Gaussian distribution centred in a frame that is lagging the LSR. In the usual $(U,V,W)$ velocity system, relative to the LSR, we have
\begin{equation} \label{eq:pdf1}
f = \frac{1}{(2\pi)^{3/2}\sigma_{_U}\sigma_{_V}\sigma_{_W}} \exp \Big[ -\frac {U^2} {2 \sigma_{_U}^2}  \\
- \frac {\left(V - V_0 \right)^2} {2 \sigma_{_V}^2} - \frac {W^2} {2 \sigma_{_W}^2} \Big]
\end{equation}
for the probability density of velocities, where $V_0$ describes the ``asymmetric drift'' --- i.e. the lag of the population centroid relative to the LSR. Most stars in the solar neighbourhood belong to the thin disk, whose kinematic characterisation we take from\footnote{Recent studies with much larger samples, and including data from the {\it Gaia\/} satellite, are broadly consistent with these figures, albeit with substantial differences arising from different selection criteria; see, e.g., \cite{2020AJ....160...43A} and \cite{2022ApJ...932...28V}.} \citet{ben03}: $V_0=-15\;{\rm km\,s^{-1}}$, $\sigma_{_U}=35\;{\rm km\,s^{-1}}$, $\sigma_{_V}=20\;{\rm km\,s^{-1}}$, and $\sigma_{_W}=16\;{\rm km\,s^{-1}}$.

With the velocity ellipsoid as given above, applying to both stars and clouds, the effective dispersion of the relative speed between clouds and stars is $\sigma_{\!e\!f\!f}\simeq 60\;{\rm km\,s^{-1}}$. Consequently the strong disruptions that arise from interactions with low values of $u_i$ ($u_i\ll 60\;{\rm km\,s^{-1}}$) occur much less often than milder disruptions at the same impact parameter. 

To check how sensitive our results are to the assumed velocity distribution of the clouds, we have also considered the limiting cases of cold disk rotation ($\sigma_{_U}=\sigma_{_V}=\sigma_{_W}=V_0=0$), and a non-rotating, spherical halo ($\sigma_{_U}=\sigma_{_V}=\sigma_{_W}=155\;{\rm km\,s^{-1}}=-V_0/\sqrt{2}$) for the \emph{cloud} distribution function (the stellar distribution function is fixed, as above). Where appropriate we give results for these two limits in addition to those for our reference case.

\subsection{The orbital plane}
\label{sec:strat}
Some of the statistics that we will construct, notably the distributions of the debris kinematics, have different contributions according to the different planes in which the cloud might orbit the star. The simulations described in Secs.~\ref{sec:repex} and \ref{sec:library} are presented in a generic Cartesian frame, with $z=0$ chosen as the orbital plane. But the labelling of the axes is, of course, unimportant, and physically the $+x$ direction in Figure \ref{fig:diagramT} is defined by the velocity of the cloud ($\boldsymbol{v}_{c}$) relative to the star ($\boldsymbol{v}_{\star}$), i.e. the direction of the vector $\boldsymbol{v}_{c}-\boldsymbol{v}_{\star}$. Thus the $(U,V,W)$ velocity components of the debris stream for an interaction with any given impact parameter, $\boldsymbol{b}$ (a two-dimensional vector, orthogonal to $\boldsymbol{v}_{c}-\boldsymbol{v}_{\star}$), can be computed from our simulation library by employing a suitable rotation of coordinates.

\begin{figure}
\begin{center}
\includegraphics[width=0.45\textwidth]{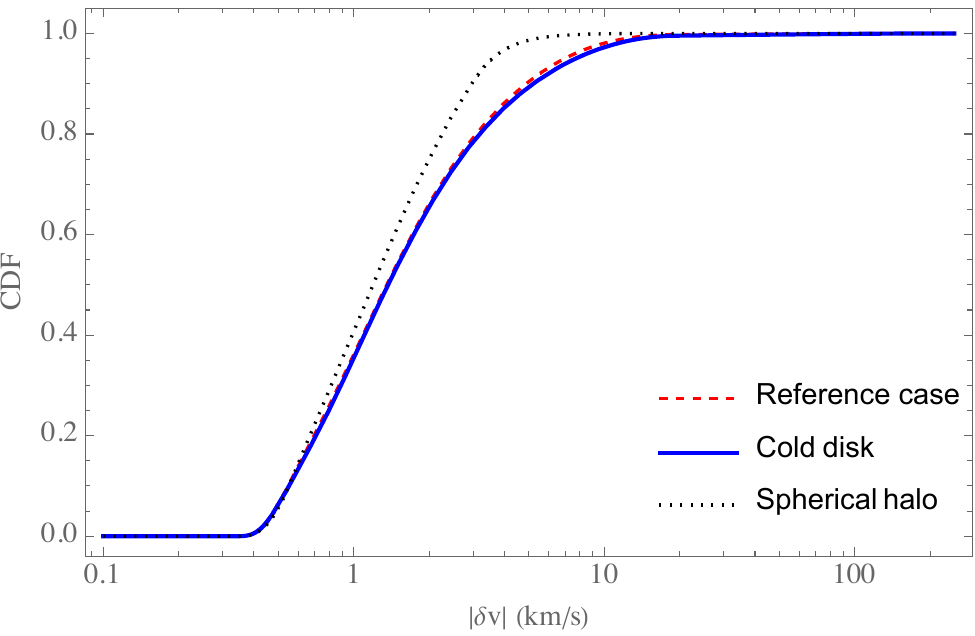}
\end{center}
\caption{Similar to Fig.~\ref{fig:machcdf}, though for the drift speed of the debris. All disruptions are substantially supersonic ($>0.18\;{\rm km\,s^{-1}}$), and the majority are hypersonic.}
\label{fig:driftCDFhydro}
\end{figure}

\subsection{Disruption rates}
\label{sec:colr}
The usual form of the rate, $\mathcal{R}$, (per unit volume, per unit time) for any two body collision process is the product of: the number densities of the two species involved in the collision; the collision cross-section; and, the relative speed. In our case, then, the differential element of rate is
\begin{equation} \label{eq:ratedifferential}
{\rm d}^8\mathcal{R} =n_\star\,n_c\;{\rm d}^2 \boldsymbol{b}\; |\bvc - \bvs|\;f(\bvs)\,{\rm d}^3 \bvs\;f(\bvc)\,{\rm d}^3 \bvc,
\end{equation}
where $n_\star$ and $n_c$ are the total number densities of stars and clouds, respectively. In \S\ref{sec:connections}, where we consider possible connections to astronomical data, it is important to know the absolute rate (and also the lifetime of the debris streams as distinct entities), for which the number densities of the colliders must be specified. However, in the rest of this section we will be concerned with probability distributions, for which the number densities factor out.

Introducing the indicator function, $\mathbf{1}({\cal C})$, which takes the value unity if the condition ${\cal C}$ is satisfied, and zero otherwise, the total disruption rate is found by integrating equation \eqref{eq:ratedifferential} over all variables:
\begin{equation} \label{eq:rateintegral}
\mathcal{R} =n_\star\,n_c\!\int\! {\rm d}^2 \boldsymbol{b}\; |\bvc - \bvs|\;f(\bvs)\,{\rm d}^3 \bvs\;f(\bvc)\,{\rm d}^3 \bvc\; \mathbf{1}({\cal C}),
\end{equation}
with ${\cal C}=\,$``Disruption'' in this case. The domain of integration corresponding to cloud disruption extends up to the red line in Figure \ref{fig:totalenergy}, and in this instance the indicator function manifests axisymmetry around the axis $\bvc - \bvs$, so we can replace ${\rm d}^2 \boldsymbol{b}$ with $2\pi b\,{\rm d}b$. We find that the integral evaluates to $\mathcal{R}_{tot}$, with 
\begin{equation} \label{eq:rate}
\frac{\mathcal{R}_{tot}}{n_\star\,n_c} \approx 2.7\times10^{-7} \qquad {\rm pc^3\,Myr^{-1}}.
\end{equation}
This figure is the total rate, but we are also interested in the probability distributions for various quantities, as described in the following sections.

For our two auxiliary cases of initial cloud velocity distribution function -- i.e. cold disk rotation, and spherical halo -- we find that the total disruption rates are $\sim 5$\%\ larger and a factor of $\sim 8$ smaller, respectively, relative to the reference case in equation \eqref{eq:rate}. The disruption rate for a spherical halo is much lower because in that case the typical relative speed between stars and clouds is very large indeed, so that most interactions are too brief to disrupt the cloud.

\subsection{Axis ratios}
\label{sec:axisr}
The cumulative probability distribution, $P(<M_{rms})$, for the axis ratio, $M_{rms}$, of the debris stream can be evaluated using expression \eqref{eq:rateintegral}, by imposing the corresponding condition (i.e. disruption, with axis ratio less than the specified value): 
\begin{equation}
P(<M_{rms}) = \frac{\mathcal{R}(<M_{rms})}{\mathcal{R}_{tot}}.
\end{equation}
As for the total rate, the integral for $\mathcal{R}(<M_{rms})$ simplifies because the indicator function is axisymmetric around the axis $\bvc - \bvs$. The resulting probability distribution is shown in Figure \ref{fig:machcdf}. As expected on the basis of Fig.~\ref{fig:machnums}, in combination with a large relative velocity dispersion, the apparent axis ratio exhibits a narrow distribution around a mean that is not much greater than $\sqrt{2}$ --- consistent with the tidal debris being predominantly flattened (``disk-like''), not elongated (``cigar-shaped''). With respect to the elongated tail of the probability distribution, discussed earlier, we find that $\approx 0.3\%$ of events in the thin-disk case have $M_{\rm rms} > 10$.

\subsection{Drift speeds}
\label{sec:driftvstat}
In much the same way as for the axis ratio, we can evaluate the cumulative probability distribution function (CDF) for the drift speed of the debris. Recall that the drift speed is a gauge of the rate at which the stream is expanding, so there is no dependence on the orientation of $\bvc - \bvs$ and the indicator function is again axisymmetric, with corresponding simplification of the rate integral. The resulting CDF is presented in Figure \ref{fig:driftCDFhydro}. There we see that the debris always expands supersonically (see Fig.~\ref{fig:massdens}), and the typical expansion speed is $\sim 2\;{\rm km\,s^{-1}}$.

\subsection{Stream kinematics}
\label{sec:kinematics}
Evaluating the full kinematic distributions for the ensemble of disrupted clouds is more complex than the CDFs for the axis ratio and drift speed, for two reasons. First, because we are dealing with vector quantities the indicator function is not symmetric around the direction defined by $\bvc-\bvs$, and consequently one must numerically evaluate each integral over all 8 of the dimensions shown in equation \eqref{eq:ratedifferential}. Secondly, because each cloud is expanding, any given debris stream is not characterised by a single velocity vector but, rather, a range of velocities around the COM value. However, Figure \ref{fig:driftCDFhydro} shows that the debris stream typically expands at a speed that is small compared to the velocity dispersion of the stars and the clouds. Consequently we expect that the bulk of the velocity distribution functions can be well approximated by neglecting the expansion of the stream and representing the debris kinematics by the motion of the COM. We adopt that approximation.

There are three velocity components to characterise; we start with the distribution of the debris streams in $U$. If the distribution is discretised into channels of width $\Delta$, then the indicator function for each channel, $U_j$, corresponds to the condition $U_j-\Delta/2<(\dot{\boldsymbol{X}}_f+\bvs)\cdot(1,0,0)\le U_j+\Delta/2$. The distributions for the $V$ and $W$ velocity components follow similarly, by contracting the debris COM velocity ($\dot{\boldsymbol{X}}_f+\bvs$) with $(0,1,0)$ and $(0,0,1)$, respectively. All three of these PDFs are very close to Gaussians with the same mean and dispersion as the corresponding distribution of the initial cloud population. In other words: the debris COM has essentially the same kinematic distribution as that of the pre-disruption clouds.

If, instead of our reference velocity distribution for the clouds, we consider clouds that are drawn from a non-rotating, spherical halo we again find that the population of debris streams exhibits essentially identical velocity distribution functions to the input cloud population. On the other hand, if the cloud population is initially in cold disk rotation (i.e. all moving with the LSR) then the debris streams manifest a floor of $\approx 6\;{\rm km\,s^{-1}}$ in their velocity dispersions, reflecting the kick that is received during interaction with the star. And for this case the tidal debris follows distributions that are not well approximated by Gaussians.

\section{Evolution of the debris stream}
\label{sec:evolution}
The evolution of tidal debris streams has been considered in detail by \cite{koch94} for the case of stars that are tidally disrupted by a massive black hole, but that circumstance is different to the one considered here. We have: no orbit self-intersections (because there is no fall-back for hyperbolic orbits); debris moving with much lower speeds; relatively benign environmental conditions; and, the stream is cold and neutral. Indeed, because we are dealing with a hydrogen snow cloud, the \htwo\ pressure is at, or close to saturation even before the cloud approaches the star, and as the tidal debris expands and cools we expect that much of the \htwo\ will condense into solid form. In Appendix \ref{sec:appa} we present a quantitative description of the adiabat appropriate to the case where \htwo\ (and possibly also He) is in phase equilibrium. From an initial temperature of $5.1\;{\rm K}$ and pressure $9.9\times 10^{-3}\,{\rm dyne \; cm^{-2}}$ (i.e. the initial conditions in the cloud specified in \S\ref{sec:hydrostatics}), we find that the gas pressure drops to the typical pressure of the ISM when $T\simeq 2\;{\rm K}$. At that point the fluid has expanded by a factor of approximately $5\times 10^{9}$, and roughly 60\% of the \htwo\ (but none of the He) has condensed. This phase of the evolution takes $\sim 7{,}000\;{\rm yr}$ and the tidal stream dimensions reach $l_1\sim l_2\sim 4\times 10^{16}\,{\rm cm}$, and $l_3\sim 4\times 10^{15}\,{\rm cm}$.

If, as is the case in some published models of \htwo-snow clouds \citep{ww19}, the fluid first saturates at higher temperatures then adiabatic expansion may condense essentially all of the hydrogen before the stream pressure drops to the typical ISM pressure --- e.g. approximately 99\%\ condensed when saturating at $24\;{\rm K}$. For simplicity, therefore, we couch our subsequent discussion in terms of just two components making up the tidal debris: gaseous helium atoms, and lumps of solid molecular hydrogen. The fate of the solid \htwo\ depends partly on the characteristic size of the individual lumps, which we proceed to estimate in the following section. For simplicity we assume that the solid \htwo\ takes the form of spherical lumps.

\subsection{Snowballs in the debris stream}
\label{sec:lumpsize}
Lumps of solid\footnote{Drops of liquid \htwo\ may also be present in the pre-disruption cloud, but for our purposes here there is no significant difference between solid and liquid forms of the condensate. The debris stream achieves very low temperatures, so any condensate that is initially in liquid form will freeze.} \htwo\ are expected to be present in the cloud before disruption, so it is not necessary to nucleate new particles and we assume that condensation proceeds mainly through growth of the pre-existing lumps. Figure 3(c) of \cite{ww19} shows \htwo\ snow making up $\sim 1$\% of the mass of the fluid in the cloud envelope, which itself constitutes $\sim$ one tenth of the total cloud mass. In that case, then, the pre-existing condensates amount to $\sim 10^{-3} M_{c}$. Thus when the disrupted cloud expands and all of the \htwo\ condenses we expect the size of the solid lumps to increase by a factor $\sim10$.

Prior to disruption there is an upper limit on the size of particles that can be maintained in suspension within the cloud: small particles are easily stirred up by convection and move with the fluid, whereas large particles settle out under the influence of gravity. For solid \htwo\ spheres of radius $a$ and density $\rho_s\approx0.087\;{\rm g\,cm^{-3}}$, in a fluid of density $\rho$ that is upwelling at speed ${\cal V}$ past a stationary particle, the drag force balances the gravitational force when
\begin{equation}\label{eq:suspension}
\frac{4}{3}\pi a^3\rho_s\, g = \pi a^2\,\rho\,{\cal V}^2,
\end{equation}
where $g$ is the acceleration due to gravity. We do not have a clear expectation for ${\cal V}$, so we simply assume that it is a modest fraction of the sound-speed $c_s$ --- specifically, we assume ${\cal V}\sim 0.1\,c_s$. The right-hand-side of equation \eqref{eq:suspension} then scales with the fluid pressure, and that varies over many orders of magnitude depending on location within the cloud. Near the cloud surface the pressure is extremely small, whereas the acceleration due to gravity is roughly constant, and there only very tiny particles can be maintained in suspension by the fluid motions. Conversely, the pressure in the very central regions of the cloud is roughly constant, but there the acceleration due to gravity becomes tiny and correspondingly the lumps of condensate can be large. 

For the purposes of the present discussion we are mainly interested in knowing the form that is taken by most of the mass of \htwo, so we adopt a mass-weighted mean, $\langle a \rangle$, of the radius, $a$ (i.e. the particle radius is averaged over the whole of the cloud, weighted by the mass of the fluid element). For our hydrostatic model (\S\ref{sec:hydrostatics}) the result is $\langle a \rangle \simeq 10\;{\rm cm}$. During the post-disruption expansion of the cloud, then, we anticipate that lumps of solid \htwo\ can grow to be as large as $a\sim 10^2\;{\rm cm}$.

\subsection{Helium in the debris stream}
\label{sec:heliumdebris}
\begin{figure}
\begin{center}
\includegraphics[width=0.45\textwidth]{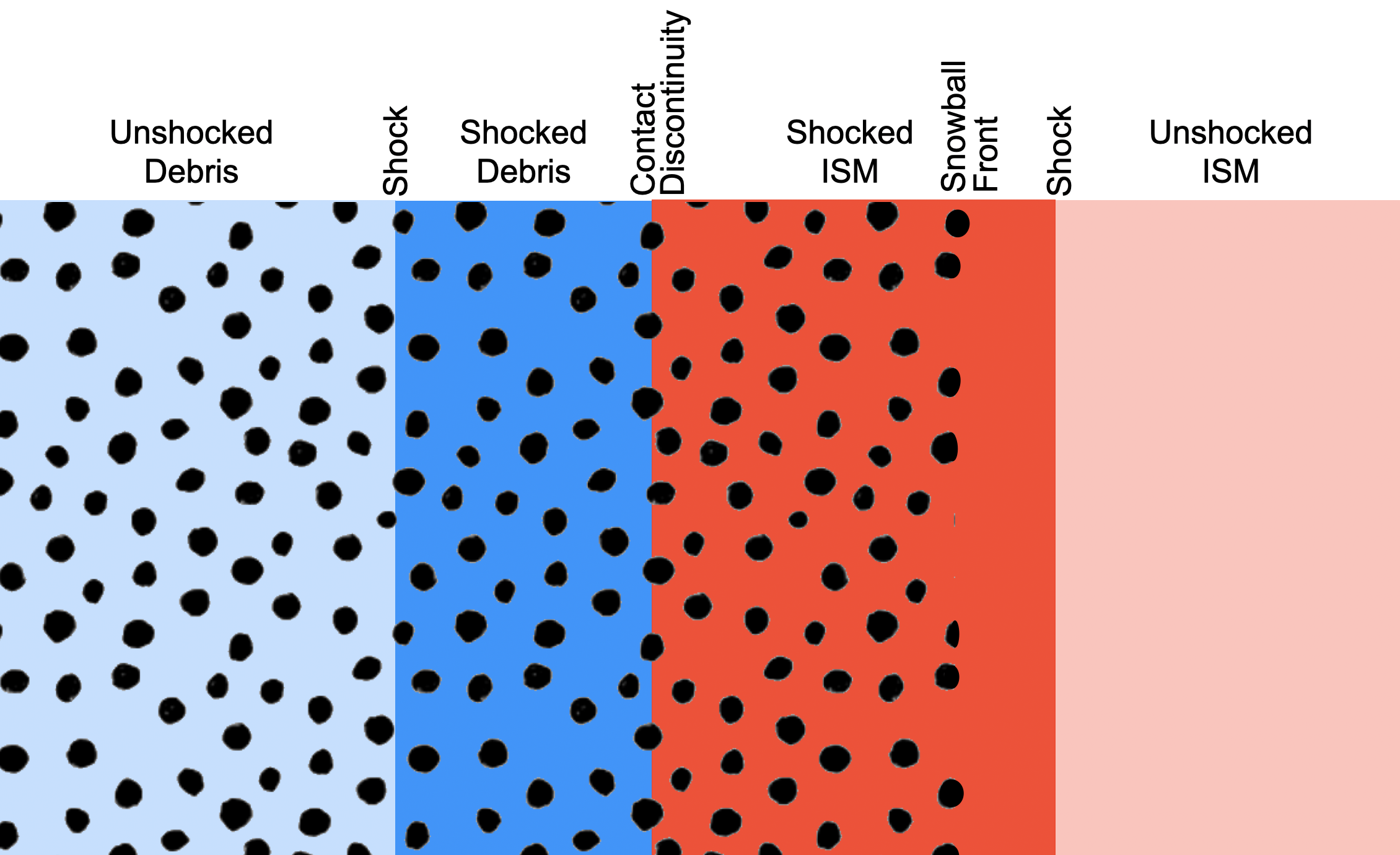}
\end{center}
\caption{Schematic of the interface between the tidal debris (shown in blue) and the ISM that it runs into (shown in red). The ISM and the helium in the tidal debris stream are both shocked, so the gas is compressed and heated as it undergoes a step change in velocity; the shocked gas is rendered in denser colours. By contrast the \htwo\ snowballs (black dots) are not directly affected by the shock in any way, so they cross the contact discontinuity and move through the region of shocked ISM.}
\label{fig:shockeddebris}
\end{figure}

In contrast to the \htwo, helium in the debris stream all remains in gaseous form and soon becomes vulnerable to outside influence --- undergoing shock heating as it ploughs into the ambient ISM. There are five distinct zones that result from this interaction, as illustrated schematically in figure \ref{fig:shockeddebris}. From left-to-right in the figure those zones are: (i) the unshocked debris stream; (ii) the shocked debris stream; (iii) the shocked ISM permeated by snowballs; (iv) the shocked ISM without snowballs; and, (v) the unshocked ISM. The snowballs are not directly affected by the gas dynamics, so where present they move with their original velocity --- i.e. the same velocity as the unshocked tidal debris. Thus in zones (ii) and (iii) the snowballs are streaming relative to the gas; and in those zones they are also evaporating, as the \htwo\ is far below its saturation pressure. 

The state of the shocked gas depends on the speed of the stream relative to the ISM that it runs into, and on the thermodynamic state of the unshocked gases. For the velocity distribution we have adopted (\S\ref{sec:priors}), the typical speed in the LSR is of order $40\;{\rm km\,s^{-1}}$. The density and temperature of the diffuse ISM can both range over orders of magnitude, according to whether the material is part of the cold, neutral medium (CNM), warm (neutral, WNM, or ionised, WIM) or hot, ionised medium (HIM) \citep[see, e.g., Table 1.3 of ][]{2011piim.book.....D} with the pressure being roughly similar in all regions. As per Appendix \ref{sec:appa}, the pre-shock conditions in the debris stream are set using the point at which the adiabatic trajectory of the fluid reaches a pressure equal to the typical pressure of the diffuse ISM. This choice is appropriate because in most cases the rate of increase of the linear dimensions of the debris stream (\S\ref{sec:driftvstat}) is greater than the speed at which the shock crosses the debris; thus, in directions away from the shock front, expansion of the stream continues until inhibited by the ambient pressure.

Conditions in the interaction region can be determined by imposing conservation of fluxes of mass, momentum and energy across the two shocks and the contact discontinuity, and then requiring that the sum of the velocity changes across the two shocks be equal to the velocity difference between the tidal debris and the diffuse ISM. The following key points apply. Pressure and velocity are the same on both sides of the contact discontinuity. The two pre-shock pressures are also equal, so both shocks propagate with the same Mach number. It follows that the speed of the tidal-debris shock is much smaller than that of the ISM shock, because of the much larger sound speed in the ISM: the ISM shock moves at $50 - 100\;{\rm km\,s^{-1}}$ (CNM -- HIM), whereas the tidal-debris shock propagates at speeds that range from $\sim 0.1\;{\rm km\,s^{-1}}$ (interaction with the HIM) up to $\sim 3\;{\rm km\,s^{-1}}$ (interaction with the CNM). Such speeds imply that the shocked tidal-debris remains neutral in all cases, with temperatures ranging from a few Kelvin up to $\sim 10^3\;{\rm K}$, whereas the shocked ISM is in all cases predominantly ionised.

As mentioned at the start of \S\ref{sec:evolution}, the stream expands to dimensions as large as $l_1, l_2\sim 4\times 10^{16}\,{\rm cm}$ before it reaches a pressure comparable to that of the diffuse ISM. The shock crossing time may be as little as $\sim 3{,}000\;{\rm yr}$, if disruption takes place in the CNM, or as long as $\sim 10^5\;{\rm yr}$ in the HIM. We note that the speed of the tidal-debris shock is comparable to the speed difference between the snowballs and the shocked gas, so the timescale on which the snowballs separate from the gaseous debris stream is comparable to the shock crossing time.

Gas in the debris stream is only slightly decelerated as it is shocked, and only declines substantially in speed after it has swept up a comparable column-density of interstellar gas (momentum conservation). In the CNM that deceleration happens on a timescale that is not very different from the shock crossing time, whereas in the warm phases of the ISM it takes much longer ($\sim 10^5\;{\rm yr}$), because of the lower density, and much longer still in the HIM ($\sim 10^7\;{\rm yr}$).

\subsection{Plasma microstructure}
\label{sec:structure}
In region (iii) the ambient ISM is compressed and shock-heated to a high temperature and a pressure exceeding that of the ambient ISM by a factor of at least a few (HIM) and possibly as much as a few thousand (CNM). This hot, high-pressure ambient can be expected to drive conductive evaporation flows from the snowballs, so each snowball leaves behind it a trail of ionised hydrogen that is cooler and denser than the ambient. The result is a vast number of parallel\footnote{The streaks of H-plasma are only locally parallel: there are velocity gradients across the unshocked debris stream, and thus the direction of relative motion of snowballs and shocked ISM must also change.} striations in electron-density, temperature, velocity and composition. 

We are not aware of any published model that allows us to estimate the properties of the snowballs' plasma trails. Conductive evaporation flows were considered in detail by \cite{1977ApJ...211..135C}, but their model relates to the evaporation of a cool, spherical plasma cloud at rest inside a hot ambient plasma. By contrast, in our case the inward heat flow must also supply the sublimation energy of the molecules from the solid surface, the binding energy of the molecules, and the ionisation energy of the atoms, in order to generate an outward flow of hydrogen plasma. Consequently we are not currently able to estimate snowball lifetimes.

\subsection{Remnant snowballs wander the Galaxy}
At present it is unclear whether there are remnant snowballs that escape from the shocked gas in the interaction region. However, those that do escape might have quite long lifetimes in the diffuse ISM; in particular we note that sublimation into a vacuum proceeds very slowly indeed once the surface of the solid has charged up \citep{2013MNRAS.434.2814W}. It is possible that conductive evaporation in the diffuse medium is the dominant erosion mechanism, particularly for snowballs that are moving supersonically through cool, dense regions of the ISM, but (as described above) at present we are not able to make a useful estimate of the rate at which that process proceeds. We do expect erosion as a result of UV photodissociation of molecules at the surface of the snowball, but in the solar neighbourhood the UV field is so weak that it cannot destroy the snowballs within the current age of the universe.

\section{Astrophysical connections}
\label{sec:connections}

In this section we consider how snow-cloud tidal debris might manifest in astrophysical data. 

\subsection{Radio-wave scattering screens}
\label{sec:scatteringscreens}
The plasma microstructure described in \S\ref{sec:structure} will scatter radio waves and is potentially of interest in connection with the scattering screens that are responsible for the intra-day variability of quasars \citep[e.g.][]{1997ApJ...490L...9K, 2000ApJ...529L..65D, 2020A&A...641L...4O} and the parabolic arcs seen in (the power spectra of) pulsar dynamic spectra \citep[e.g.][]{2001ApJ...549L..97S, 2023MNRAS.518.1086M}. The points of interest are as follows. First, depending on the phase of the ambient ISM, the microstructure can have electron densities that are much greater than the values deduced for the WIM -- a property that is also inferred for the radio-wave scattering screens \citep{2011AIPC.1366..107R, 2013MNRAS.429.2562T}. Plasma densities in the snowball trails are greatest in cases where the debris stream is interacting with the CNM.

Secondly, the snowballs are streaming relative to the shocked gas, and thus the plasma microstructure will be elongated along the direction of relative motion. The resulting scattering of radio-waves will be anisotropic to the same degree. This is interesting because anisotropy is often observed in radio-wave scattering \citep[e.g.][]{2008MNRAS.388.1214W, 2009MNRAS.397..447W, 2019MNRAS.487.4372B, 2022MNRAS.513.2770B, 2022MNRAS.515.6198S}. We caution that it is unclear whether the high levels of anisotropy that are often observed can arise in this way, as the streaming velocities are limited to at most a few ${\rm km\,s^{-1}}$; a detailed model of the evaporation flow is needed to address this question.

Finally, snow-cloud tidal debris is tiny by the usual standards of the ISM, with major axes growing to only $\sim 3{,}000\;{\rm AU}$, and that is at least qualitatively consistent with the transient nature of the intra-day variability phenomenon in some sources \citep{2006MNRAS.369..449K, 2008ApJ...689..108L, 2015A&A...574A.125D}. To date, though, observations have not been able to furnish us with firm indications of the size of the scattering screens so a quantitative comparison is not yet possible.

In contrast to the aspects listed above, the ``disk-like'' shapes that the present model predicts to be typical for tidal debris is at odds with the filamentary scattering screen morphology reported by \cite{wang21}. And as that is the only instance where a screen morphology has been established, the disagreement is a serious one. This is an ironic situation, because \cite{wang21} suggested that the observed morphology might be well explained by the tidal disruption of a cold cloud, and that suggestion formed part of the motivation for the present study. The reason for this difference in outlook is the following. In the present paper we have concerned ourselves with TDEs of clouds by unassociated stars, so the orbits are hyperbolic. On the other hand the perspective of \cite{wang21} was based on TDE simulations in the literature \citep[e.g.][]{koch94}, and they are almost exclusively for parabolic orbits. We return to the issue of stream morphology in \S\ref{sec:discussion}, where we consider the circumstance of clouds that are loosely bound to ``host'' stars. In that case the clouds can be tidally disrupted by their hosts, and near periastron the orbits of disrupting clouds would be well approximated by parabolae.

More serious than the morphological mismatch described above is the fact that the present model cannot explain the observed number density of scattering screens -- as discussed in \S\ref{sec:localdensity} -- even under the generous assumption that all of the local dark matter is composed of snow clouds.

\subsection{Scattered light, IR cirrus, and atomic hydrogen}
It is possible that snowballs could be long-lived entities. In that case they will follow orbits through the Galaxy that are dictated by their initial conditions (position and velocity) and the Galactic potential. These itinerant snowballs should contribute at some level to a diverse set of astrophysical phenomena: scattering of photons (optical/IR, or hard X-rays) from background sources; far-IR and mm-wave thermal emission; and, each snowball will leave a trail of atomic hydrogen, as a result of UV-photodissociation of molecules at the surface of the snowball. We note, though, that the low density of debris streams in the solar neighbourhood (see below) means that the disruption process studied in this paper cannot contribute much to the observed 21cm signal.

\subsection{The density of debris streams in the local ISM}
\label{sec:localdensity}
To estimate the density of snow-cloud tidal debris streams in the solar neighbourhood we first need to establish the absolute normalisation of the disruption rate \eqref{eq:rate}. The density of stars in the Galactic plane near the Sun is $\approx 0.137\;{\rm pc^{-3}}$ \citep{kroup93}, following an exponential distribution with scale-height $\approx 300\;{\rm pc}$ as one moves away from the plane. However, most of these stars have masses lower than the fiducial value ($1\;{\rm M_\odot}$) we have used for our calculations. The appropriate weighting for stars of different masses is demonstrated by equation \ref{eq:disruptionboundary}: the disruption cross-section is proportional to the stellar mass. With this weighting applied to the stellar mass function \citep[][their equation 13]{kroup93}, the effective number-density of stars in the solar neighbourhood decreases to $n_\star\simeq 0.064\;{\rm pc^{-3}}$.
For the initial population of clouds, on the other hand, we adopt the local dark density ($\rho\approx 0.013\;{\rm M_\odot\,pc^{-3}}$) inferred by \cite{mck15} which, for $M_{c} = 3\times 10^{-5}\,{\rm M_\odot}$ as adopted here, corresponds to\footnote{This value is actually an upper limit, because the local dark density estimated by \cite{mck15} includes all forms of dark matter, not just snow clouds.} $n_c\approx 400\;{\rm pc^{-3}}$. Furthermore, because we assume that the clouds have the same kinematic distribution as the stars we also assume that the cloud number density declines exponentially with the same scale-height as the stars. These assumptions yield a total TDE rate in the solar neighbourhood of
\begin{equation} \label{eq:localrate}
\mathcal{R}_{tot} \approx 7\times10^{-6} \,\exp(-|Z|/h) \qquad {\rm pc^{-3}\,Myr^{-1}},
\end{equation}
where $Z$ is the height above the Galactic plane, and the scale-height on which the rate declines is $h=150\;{\rm pc}$.

At present we have no reliable estimate for the lifetime of the microstructured plasma, so we simply assume it is no greater than the deceleration timescale of the stream in the HIM --- i.e. $\lta 10^7\;{\rm yr}$ (\S\ref{sec:heliumdebris}). With that assumption, equation \eqref{eq:localrate} implies a number density of debris streams of $\lta 7\times 10^{-5}\;{\rm pc^{-3}}$ in the neighbourhood of the Sun (where $|Z| \ll h$).

With the exception of morphology, there is a similarity between the expected properties of {\it individual\/} tidal debris streams and the inferred properties of radio-wave scattering screens --- as described in \S\ref{sec:scatteringscreens}.  However, the following argument demonstrates that the model we have described cannot explain the observed population of radio-wave scattering screens. With an assumed size of $\sim 4\times 10^{16}\;{\rm cm}$ for the tidal debris, the line of sight to a source that is $\sim 1\;{\rm kpc}$ distant is expected to intersect a total number of $\sim 4\times 10^{-5}$ streams. By contrast, pulsar dynamic spectra indicate $\sim 1$ strongly scattering screen for most sources, out to $\sim 1\;{\rm kpc}$ \citep{2006ChJAS...6b.233P, 2022ApJ...941...34S}, and the true density might be as much as two orders of magnitude higher when weakly-scattering screens are counted \citep{2024arXiv241021390R}.

\vskip 1cm

\section{Discussion}
\label{sec:discussion}
The low rate that we estimate for snow-cloud TDEs stems, in part, from the small cross-section for that process, of order $10^2\;{\rm AU^2}$ for relative speeds of $\sim 30\;{\rm km/s}$ (see Fig.~\ref{fig:totalenergy}), coupled with the assumption that there is no physical relationship between the cloud and the star that disrupts it. If, instead, we were to assume that clouds are loosely gravitationally bound to stars, then the TDE rate \eqref{eq:localrate} may be much larger. If clouds are orbiting at a large distance from the host star then -- just as with Oort's model for long-period comets \citep{1950BAN....11...91O} -- only a small impulse is required, e.g. from a passing star, to shift any one of them onto a new, plunging orbit that leads to destruction by the tides of the host star. And because a small impulse suffices, the perturbing star need not pass very close to the cloud; hence the cross-section for this process can be orders of magnitude larger than for unassociated populations. If so, tidal debris streams would be present in much greater numbers than we estimated in  \S\ref{sec:localdensity}. Furthermore, the disrupting orbits in this case should be close to parabolic, and therefore the tidal debris streams are expected to be highly elongated \citep[e.g.][]{koch94}. Thus the hypothesis that clouds are bound to, and disrupted by, ``host'' stars appears to remove the two key objections to a disrupted snow-cloud interpretation of radio-wave scattering screens, while retaining all the attractive features described in \S\ref{sec:scatteringscreens}. This interesting possibility will be the subject of a future paper.

The affine formalism that we have employed permits the fluid only a small number of degrees of freedom, so there is plenty of motivation for future investigations of cold-cloud TDEs in full hydrodynamic detail and making use of powerful computers. We anticipate that partial disruptions may be a common feature of such computations, as realistic models of snow clouds exhibit a high density core surrounded by an extended, low-density envelope \citep{ww19}.

One aspect of snow-cloud TDEs that merits particular attention with a more detailed hydrodynamic model is the possibility that \htwo\ snowballs could aggregate into much larger lumps, under the influence of gravity. It has previously been suggested, in the context of stars (on parabolic orbits) disrupted by massive black holes, that self-gravity of the debris stream can be important \citep[e.g.][]{koch94, 2020ApJ...900L..39C}. At first sight that possibility might seem unpromising, precisely because the gravitational field of the disruptor has overwhelmed that of the disrupted body. However, disruptions are anisotropic and for some TDEs the stream expands freely along its major axis while at the same time self-gravity is important in the dynamics of the minor axes \citep{koch94}. For snow cloud TDEs self-gravity is especially interesting because the vast number of lumps of solid \htwo\ that are present in the debris stream constitute a ``gas'' of particles that has density but not pressure, and that ``gas'' will therefore be prone to gravitational growth of density inhomogeneities. If any solid \htwo\ aggregates of size $\ga 10^2\;{\rm m}$ form in the debris stream, under the influence of gravity, then snow-cloud TDEs could be a source of interstellar objects such as 1I/`Oumuamua \citep{2017Natur.552..378M}. Indeed, as noted in the Introduction, it has already been proposed that 1I/`Oumuamua might be principally composed of solid \htwo\ \citep{2018A&A...613A..64F, 2020ApJ...896L...8S}, but understanding the origin of such a large lump of \htwo\ is challenging \citep{lev21}.

\section{Summary and Conclusions}
\label{sec:conclusions}
We have investigated the process of tidal disruption of hydrogen snow clouds by stars in the case where there is no prior physical association between the two. Our approach employs a restricted hydrodynamics scheme formulated by Carter and Luminet, in which the cloud assumes the form of a triaxial ellipsoid at all times --- starting off as a sphere, but ending up with very different axis ratios in instances where disruption occurs. The formalism cannot be used to study morphologically complex features such as fallback accretion, tidal stripping or partial disruptions. However, it has a low computational cost and thus it is well suited to our task of exploring a new parameter space via simulations of a large number of different star-cloud interactions. Also for the sake of low computational cost, we adopted a polytropic EOS with index $n=3/2$, appropriate to adiabatic convection in an ideal, effectively monatomic gas. We restricted attention to a single mass ($3\times 10^{-5}\,{\rm M}_\odot$), radius ($0.78\;{\rm AU}$) combination for the cloud, and the star was assumed to be of one solar mass. For this pair we built a library of simulations describing collisions spanning a range of impact parameters and relative speeds, using an energy criterion to identify the parameter combinations that lead to TDEs.

From the model collision library we determined the statistical properties of the disruptions -- event rates, and the kinematics and morphology of the tidal debris -- using identical model velocity distribution functions for both clouds and stars (appropriate to a thin, Galactic disk population in both cases). We found that tidal debris is typically flattened (``disk-like''), in the case we investigated where the orbits are hyperbolic. We also found that the stream velocities are often quite high (tens of ${\rm km\,s^{-1}}$), as seen in the LSR. However, that is mainly a reflection of the assumed initial velocity distribution of the clouds --- as we illustrated with computations for two additional cloud velocity distribution functions.

Perhaps the most interesting aspect of our results relates to the composition of the debris stream, which is born as a gas of cold atomic helium peppered with molecular hydrogen snowballs up to about a metre in size. (Either of these components could also host metals, if they are present in the pre-disruption cloud, but we did not consider metals in this paper.) Subsequently the helium undergoes shock heating as the tidal debris ploughs through the diffuse ISM. The hydrogen snowballs are not directly affected by the shock, but are gradually eroded as they stream through the shocked gas. Snowballs in the shocked ISM leave tiny trails of ionised hydrogen that constitute plasma microstructure, and  these microstructured regions are somewhat reminiscent of the ``scattering screens'' that cause scintillation in compact radio sources.

It is possible that remnant snowballs could be long-lived, and in that case they would contribute to light scattering, to the infrared cirrus and (via UV photodissociation of the surface layers) to the 21cm signal. Much larger lumps of solid \htwo\ could, in principle, arise during snow-cloud TDEs, as a result of the mutual gravitational attraction of the snowballs, but our simplified hydrodynamic scheme does not allow us to address that possibility.

Although the properties of the tidal debris are of interest in connection with known astrophysical phenomena, disruptions are rare for the model investigated here --- in which TDEs are the result of close encounters between a star and an unrelated snow-cloud. Consequently the expected number-density of debris streams in the Galaxy is too low to be able to explain the observed radio-wave scattering screens. However, if most snow clouds are, instead, loosely bound to host stars then they can be disrupted by the host star itself. In that case disruptions are precipitated by a cloud's orbit being perturbed into the loss cone -- as with Oort's model for long-period comets -- for which a distant encounter with a field star suffices, and thus the disruption rate may be orders of magnitude higher.

\section*{Acknowledgements}
We thank Artem Tuntsov for pointing out an error in the original manuscript, and for helpful input on both the physics and the manuscript itself. We are also indebted to the referee for correcting a misunderstanding in our treatment of the shock-heating of the tidal debris. AGS acknowledges support provided by the Conselleria d'Educaci{\'o}, Cultura, Universitats i Ocupaci{\'o} de la Generalitat Valenciana through Prometeo Project CIPROM/2022/13 for parts of this work that were completed while in Spain.

\section*{Data availability statement}
Observational data used in this paper are quoted from the
cited works. Additional data generated from computations
can be made available upon reasonable request.



\appendix

\section{Adiabatic expansion with condensation of \htwo\ and Helium} \label{sec:appa}
The adiabatic expansion of an ideal gas follows a familiar trajectory. At low temperatures the vibrational and rotational degrees of freedom of the \htwo\ are not excited, so the gas is effectively monatomic, the ratio of specific heats is $5/3$, and the pressure-temperature relation is $P\propto T^{5/2}$. The tidal debris stream may initially follow this trajectory as it expands. However, as the  expansion continues the vapour pressure of \htwo\ reaches the point of saturation, and thereafter the adiabatic trajectory changes dramatically because each molecule of \htwo\ that condenses releases latent heat $b \gg kT$.

Following \cite{ww19} we can construct a simple approximate form for the condensing adiabat by neglecting the entropy and volume of the condensate in comparison with that of the gas, and modelling the saturated gas as an ideal gas at the saturation pressure. At temperatures well below the critical temperature the saturation pressure is well approximated by the following form:
\begin{equation}
P_{sat}(T)=kT{{(2\pi \mu kT)^{3/2}}\over{h^3}}\exp\left(-{{b}\over{kT}}\right),
\end{equation}
with $h$ and $k$ being Planck's and Boltzmann's Constants, respectively, and $\mu$ the particle mass.
The Sackur-Tetrode entropy per particle (expressed in units of $k$) of the saturated gas is then
\begin{equation} \label{eq:stentr}
\chi = {5\over 2} + {b\over{kT}}
\end{equation}
\citep[see sections 2.4 and 2.5 of][]{ww19}. For \htwo\ the accuracy of these approximations is demonstrated by the comparison with experimental data in Appendix B of \cite{ww19}.

Bearing in mind that the total number of particles of saturated gas, $N_{sat}$, depends on both the temperature and volume ($V$) of the fluid parcel under consideration, the total entropy increment of the saturated component evaluates to
\begin{equation}
{\rm d}S_{sat} = N_{sat}k\,\left[\chi\, {\rm d}\log\,V  + \left(\chi^2 - 2\chi + {5\over 2} \right)\, {\rm d}\log\,T \right].
\end{equation}
And, as usual for an ideal gas, the total entropy increment of the unsaturated component is just
\begin{equation} \label{eq:unsat}
{\rm d}S_{un} = N_{un}k\,\left[{\rm d}\log\,V  + {3\over 2}\, {\rm d}\log\,T \right].
\end{equation}

There are thus three regimes to consider. First, if both gases are unsaturated then we have ${\rm d}S = {\rm d}S_{un}$, which is zero for the adiabat and equation \eqref{eq:unsat} then yields the familiar result ${\rm d}\log\,T = -(2/3)\, {\rm d}\log\,V$.

When the \htwo\ reaches its saturation pressure we enter the second regime where $0={\rm d}S = {\rm d}S_{un} + {\rm d}S_{sat}$, which corresponds to an adiabatic temperature derivative:
\begin{equation}
-{{{\rm d}\log\,T }\over {{\rm d}\log\,V}} =  {{N_{un}+\chi\,N_{sat}}\over{{3\over2}N_{un}+\left[\chi(\chi-2) +{5\over 2}\right]\,N_{sat}}}.
\end{equation}
Here $N_{un}$ is just the (fixed) total number of helium atoms in the fluid parcel under consideration, $N_{sat}$, is the total number of gaseous hydrogen molecules in that parcel, and $\chi$ relates to the saturated \htwo\ so the corresponding value of $b$ in equation \eqref{eq:stentr} is the sublimation energy of \htwo. The total number of gas molecules in the fluid parcel can be rewritten as $N_{sat}=V\,n_{sat}(T)= VP_{sat}(T)/kT$. And if the \htwo\ first saturates at temperature $T_o$ and volume $V_o$ (i.e. there is zero condensate at that point), then the total number of helium atoms can be written as $N_{un}=y\,V_o\, n_{sat}(T_o)$, where $y$ is the helium abundance (by number, relative to hydrogen molecules).

\begin{figure}
\begin{center}
{\includegraphics[width=0.45\textwidth]{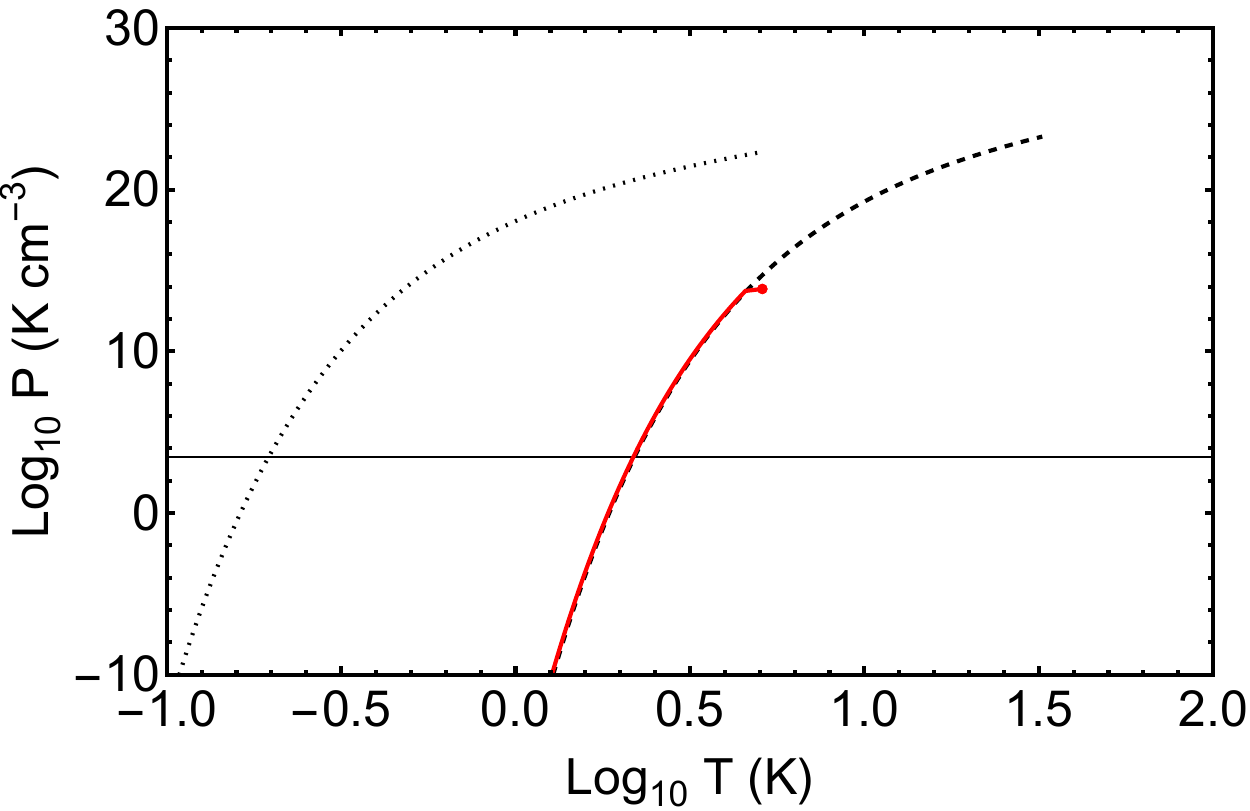}\hskip1cm\includegraphics[width=0.45\textwidth]{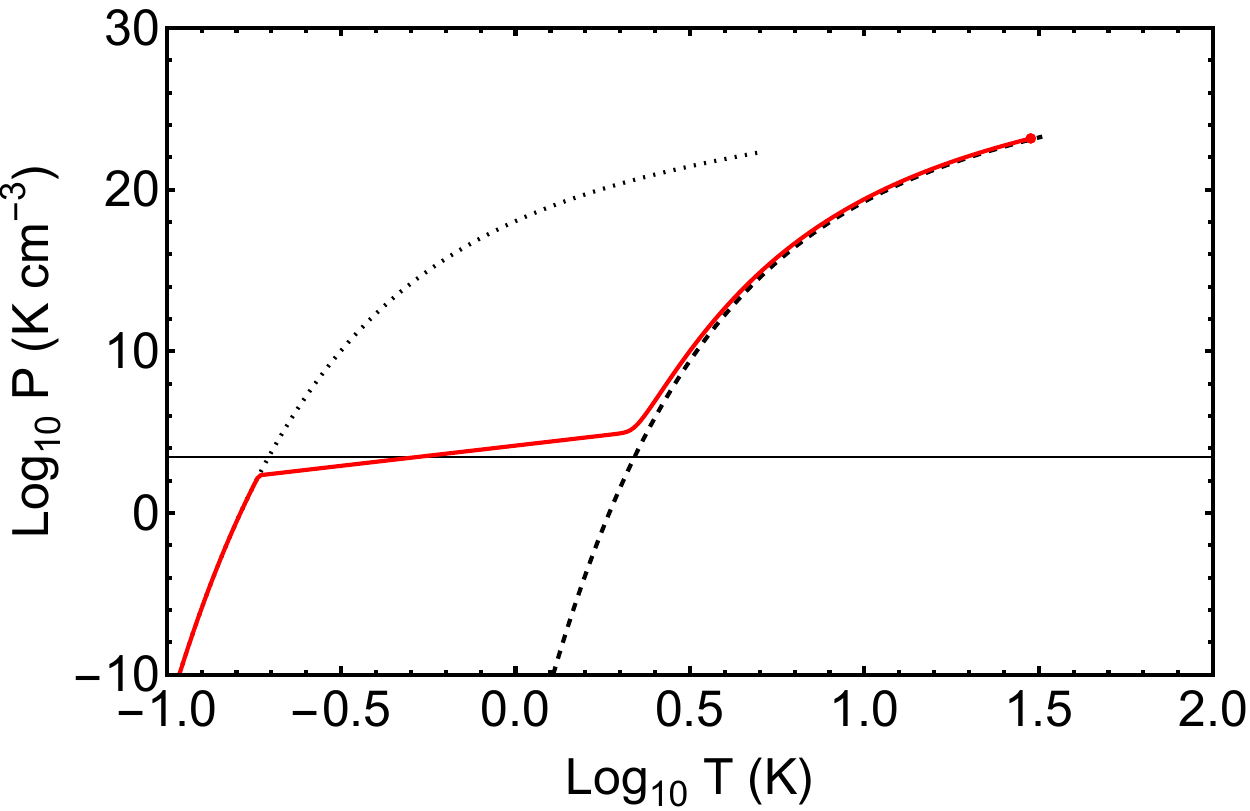}}
\end{center}
\caption{The trajectories of adiabatic expansion (red lines) in the Temperature-Pressure plane, for fluid parcels that first reach  \htwo\ saturation at $T_o\simeq4.56\;{\rm K}$ (left panel), and $T_o = 30\;{\rm K}$ (right panel). In both cases the helium to \htwo\ ratio by number is $y=1/6$. Also shown in both panels are the saturation pressures for \htwo\ (dashed curves), and He (dotted curves); and, a notional interstellar pressure of $3{,}000\;{\rm K\,cm^{-3}}$ (thin black line). The left panel is appropriate to fluid in the conditions typical of the interior of the gas cloud specified in \S\ref{sec:hydrostatics} of this paper. For comparison we have also constructed (right panel) the adiabat for a fluid parcel that first saturates at a much higher temperature --- only slightly below the critical temperature for \htwo.}
\label{fig:adiabats}
\end{figure}

Finally there is a regime where helium also saturates; this regime was not considered by \cite{ww19}. In this case the entropy interval is formally the sum of two saturated gas contributions, but because the saturation pressure of helium is so much larger than that of \htwo\ (see Figure \ref{fig:adiabats}) it is typically a good approximation to neglect the contribution of hydrogen. In that case the temperature derivative on the adiabat is just (${\rm d} S_{sat} = 0$)
\begin{equation} \label{eq:heliumtemp}
-{{{\rm d}\log\,T }\over {{\rm d}\log\,V}} =  {{2\chi}\over{2\chi(\chi-2) + 5}},
\end{equation}
with $\chi$ (and corresponding latent heat $b$) here being that of helium. Equation \eqref{eq:heliumtemp} makes clear that adiabatic volume changes of a saturated gas involve very little temperature change: the right hand side is $\sim  1/\chi = {kT} / b \ll 1$. Whereas adiabatic $P\,{\rm d}V$ work in an ideal gas can only flow to-or-from the thermal energy of the gas particles, in a saturated gas the $P\,{\rm d}V$ work also flows to-or-from the large latent heat reservoir associated with the phase change, and only a small temperature change results.

With the equations given above we can construct the full trajectory for an adiabatic expansion. To proceed we assume a helium-to-hydrogen ratio of $y=1/6$ (i.e. helium is 25\%\ of the total mass). For \htwo\ the critical temperature is $T_{crit}\simeq 32.9\;{\rm K}$ and the latent heat is $b/k\simeq 91.5\;{\rm K}$ \citep[][and references therein]{ww19}. For He, on the other hand, $T_{crit}\simeq 5.2\;{\rm K}$ and $b/k\simeq 7.2\;{\rm K}$ \citep{1998JPCRD..27.1217D}. Figure \ref{fig:adiabats} (left panel) shows the adiabat for a fluid parcel that starts with the thermodynamic conditions specified in  \S\ref{sec:hydrostatics} for the interior of our model cloud; also shown for comparison (right panel) is the adiabat for a fluid parcel that first saturates near the critical temperature for \htwo.  The left panel shows only the first and second regimes of behaviour discussed above, whereas the right panel illustrates the second and third regimes. However, the adiabat in the right panel does also exhibit a region [$0.2 \lta T ({\rm K}) \lta 2$] where there is so little gas-phase \htwo\ that the thermodynamics are controlled by the (unsaturated) helium and thus the behaviour is close to that of an ideal gas.

Regarding the discussion provided in \S\ref{sec:evolution}, the following properties of the two adiabats shown in Fig.~\ref{fig:adiabats} are of interest. For the fluid parcel shown in the left panel: $P=P_{ism}$ at $T\simeq 2.2\;{\rm K}$ and $V\simeq 5\times 10^9$ (in units of the initial volume); the gas density is $6\times 10^{-21}\,{\rm g\,cm^{-3}}$; and, at this temperature 57\% of the \htwo\ has condensed. For the fluid parcel shown in the right panel: $P=P_{ism}$ at $T\simeq 0.53\;{\rm K}$, $V\simeq 1.2\times 10^{17}$; total gas density is $4\times 10^{-20}\,{\rm g\,cm^{-3}}$, and all but a fraction $10^{-59}$ of the \htwo\ has condensed. In neither case does helium saturate before the pressure falls to the typical ISM pressure.

\label{lastpage}

\end{document}